\documentclass[pre,aps,10pt,twocolumn,floatfix]{revtex4-1}

\usepackage{geometry}
\usepackage{graphicx}
\usepackage{amsfonts}
\usepackage{amsmath}
\usepackage{bm}
\usepackage{url}
\usepackage{color}
\usepackage{xfrac}

\newcommand{\ev}{{\bm e}}
\newcommand{\la}[1]{\left#1}
\newcommand{\ra}[1]{\right#1}
\newcommand{\e}{\epsilon}
\newcommand{\av}{{\bm a}}
\renewcommand{\v}{\bm}
\def\r{\right}
\def\l{\left}

\begin{document}
\title{Emergence of limit-periodic order in tiling models} 

\author{Catherine Marcoux}
\affiliation{Physics Department, Duke University, Durham, NC 27708}

\author{Travis W. Byington}
\affiliation{726 W Main St., Madison, WI 53715}

\author{Zongjin Qian}
\affiliation{University of Chicago Booth School of Business, 5807 S. Woodlawn Avenue, Chicago, IL 60637}

\author{Patrick Charbonneau}
\affiliation{Chemistry Department, Duke University, Durham, NC 27708}
\affiliation{Physics Department, Duke University, Durham, NC 27708}

\author{Joshua E.~S.~Socolar}
\email[]{socolar@phy.duke.edu}
\affiliation{Physics {D}epartment, Duke University, Durham, NC 27708}

\date{\today}

\begin{abstract}
A 2D lattice model defined on a triangular lattice with nearest- and next-nearest-neighbor interactions based on the Taylor-Socolar monotile is known to have a limit-periodic ground state.  The system reaches that state during a slow quench through an infinite sequence of phase transitions.  We study the model as a function of the strength of the  next-nearest-neighbor interactions, and introduce closely related 3D models with only nearest-neighbor interactions that exhibit limit-periodic phases.  For models with no next-nearest-neighbor interactions of the Taylor-Socolar type, there is a large degenerate classes of ground states, including crystalline patterns and limit-periodic ones, but a slow quench still yields the limit-periodic state.  For the Taylor-Socolar lattice model, we present calculations of the diffraction pattern for a particular decoration of the tile that permits exact expressions for the amplitudes, and identify domain walls that slow the relaxation times in the ordered phases.  For one of the 3D models, we show that the phase transitions are first order, with equilibrium structures that can be more complex than in the 2D case, and we include a proof of aperiodicity for a geometrically simple tile with only nearest-neighbor matching rules.  
\end{abstract}


\maketitle 

%
\section{Introduction} \label{sec:intro}

The possibility of spontaneous formation of translationally ordered, nonperiodic structures has garnered much attention in both the physics and tiling theory communities since the discovery of quasicrystals in the early 1980’s.~\cite{Shechtman1984,Levine1984}  Quasicrystals combine long-range translational order with point group symmetries that are incompatible with periodic structure.  One conceptually fruitful approach to understanding the stability of quasicrystalline alloys has been to describe their atomic structure as a decoration of a small number of unit cell types, or prototiles, that are then arranged to form a quasiperiodic, space-filling tiling.~\cite{SteinhardtBook}  In such models, the interactions between tiles are represented by {\em matching rules} that determine which local configurations have low energy. One can show that any tiling that satisfies the matching rules everywhere must be quasiperiodic.~\cite{Penrose1974,Socolar1986,Katz1988,Levitov1988,Socolar1990,Kalugin2005}  Matching rules are also known to exist for all tilings in two or more dimensions that are generated by substitution rules~\cite{GoodmanStrauss1998,Fernique2010}, which can produce more exotic types of long range order.

One type of nonperiodic tiling that can be forced by local matching rules has been known since the discovery by Berger of a nonperiodic set of Wang tiles.~\cite{Berger1966}, later refined substantially by Robinson~\cite{Robinson1971}, and reduced to a two-tile set by Goodman-Strauss~\cite{GoodmanStrauss1999a}.  
These prototiles endowed with matching rules admit no periodic space-filling tilings, but do allow the plane to be covered in a pattern that consists of a union of an infinite set of periodic structures of ever increasing lattice constant.  Such patterns are termed {\em limit-periodic} and have point group symmetries compatible with periodicity, but no smallest reciprocal space lattice vector.

For limit-periodic tiling models, one can typically view the structure as a periodic array of decorated tiles (squares, in the case of Wang tiles), with the limit-periodic structure being displayed in the pattern of orientations of the decorations.  Denoting each orientation of each tile type as a distinct ``spin,'' one can express the matching rules that enforce the limit-periodic structure as a Hamiltonian governing local spin configurations.  The resulting spin model on a lattice can then be studied at finite temperature.  For a version of the Robinson tiles, Mi\c{e}kisz showed that one expects an infinite set of distinct thermodynamic equilibrium phases as temperature is lowered.~\cite{Miekisz1990}.  

Socolar and Taylor recently introduced a hexagonal prototile with a decoration that forces limit-periodic tilings.~\cite{SocolarTaylor2010a}.  Most remarkably, the tiling requires only a single prototile (together with its mirror image), though the rules for the 2D version include constraints on relative orientations of next-nearest-neighbor tiles. They also showed that a 3D version of the prototile could be designed so that the matching rules are enforced purely by the shape of the tile and the space-filling constraint; i.e., with no non-contact interactions.~\cite{SocolarTaylor2010b}  However, the shape of the 3D tile is highly nontrivial and the spontaneous assembly of the structure extremely hard to envision.  

Key thermodynamic properties of a lattice model based on the 2D Socolar-Taylor tile were reported by Byington and Socolar, who presented strong evidence that the system undergoes an infinite sequence of phase transitions if cooled sufficiently slowly, identified a set of order parameters for the transitions, and found approximate (but highly accurate) scaling relations between the values of the order parameters in equilibrium at rescaled temperatures.~\cite{Byington2012}  In the present paper, we review those results and extend them in several directions.  Our primary interest is in determining how the rules that require a complex tile shape might be relaxed without losing the thermodynamic stability and dynamical accessibility of the limit-periodic ground state.

We present three primary new findings.  
First, a modification of the Taylor-Socolar tile allows for a 3D face-centered cubic (FCC) lattice model with only nearest-neighbor interactions that still permit only limit-periodic ground states.
Second, the Hamiltonian for the 2D Taylor-Socolar model can be simplified substantially while retaining the path to limit-periodicity through an infinite sequence of phase transitions, even though the simplified model admits periodic ground states as well as limit-periodic ones.
Third, a similar simplification of the 3D model shows similar behavior, but the transitions in this case are first order.
In addition, this paper presents results on two aspects of the 2D tiling models that may be relevant for interpreting experiments on systems that embody the interactions necessary for producing the Taylor-Socolar structure: (1) diffraction patterns and (2) domain wall dynamics.  
We display the diffraction patterns for special decorations of the 2D models that allow particularly efficient calculations, and we identify certain types of domain walls that are highly stable.

The path to limit-periodicity in the simplified model is a surprising result, particularly if one allows for the possibility that weak next-nearest neighbor interactions might favor the periodic phase at $T=0$.  It shows that a system with a periodic ground state can self-assemble into a perfectly ordered, nonperiodic state at $T=0$ through a quasistatic process.  The key here is that at any finite temperature the entropy of the partially ordered limit-periodic phase favors that phase over any kinetically accessible periodic phase that might compete with it.  At each stage in the hierarchy of transitions, the transition to the relevant periodic phase is preempted by a transition to a partially ordered phase that is incompatible with the periodic one.  We note that this scenario is quite different from the entropic stabilization of long-range order in random tiling models for quasicrystals, which are expected to undergo transitions to crystalline states at low temperatures.~\cite{Shaw1991,Widom1993}  In the present case, there is no extensive entropy in the limit-periodic state reached at $T=0$, though entropic effects play a crucial role in guiding the system to this state in a slow quench.  

The paper is organized as follows.  In Section~\ref{sec:defns}, we give more precise definitions of our terms and of the four distinct models that we study.  In Section~\ref{sec:2Dmodels}, we review the results of Ref.~\cite{Byington2012} and present results on the behavior of the Taylor-Socolar model as a function of the strength of the next-nearest-neighbor interactions, including the special case where the next-nearest-neighbor interaction is completely absent.  The latter case represents a substantial simplification of the hexagonal prototile, both because the matching rules can easily be enforced by pairwise interactions between adjacent tiles only, and because the tile is no longer chiral, so a racemic mixture is no longer needed.  From a tiling theory perspective, this prototile is not of great interest as it admits periodic tilings.  The physics, however,  is surprising: only the limit-periodic phase forms upon slow quenching.

In Section~\ref{sec:3Dmodels}, we consider a different approach to the simplification of the tiling model.  Here we present a rhombohedral prototile that has the shape of the unit cell of a 3D FCC lattice.  The next-nearest-neighbor matching rule for the Taylor-Socolar tile is now implemented as a nearest-neighbor rule in a hexagonal layer normal to the $111$ direction in the FCC lattice.  The nearest-neighbor rule for the Taylor-Socolar tile must be weakened a bit, however, in a manner explained below.  We prove in the Appendix that this prototile does indeed admit only limit-periodic structures.  We also present a careful study of the thermodynamics of this model for the case where the in-plane rules (analogous to the next-nearest-neighbor rules of the 2D model) are absent.  This model is shown to display a highly complex set of ground states, some of which are periodic, and to reach one of the limit-periodic states upon slow quenching via a sequence of first order transitions.  Sections~\ref{sec:barriers} and~\ref{sec:diffraction} present our results on domain wall effects in the kinetics of equilibration and on diffraction patterns, respectively.  We close with a brief summary and some remarks on open questions.

%
\section{Definitions} \label{sec:defns}
The lattice models treated in this paper are derived from tiling models, which we define as follows.
\begin{itemize}
\item A \emph{tile} is a closed, compact set of points in $\mathbb{R}^n$ with an assigned integer $i$ indicating its \emph{type}. 
\item A \emph{tiling} is a set of tiles that collectively cover the entire space $\mathbb{R}^n$ with no two tiles sharing any interior points.  (Adjacent tiles share only points on their boundaries.)
\item A set of \emph{matching rules} for a tiling is a specification of allowed configurations of pairs of tiles; i.e., a specification of which tile types are allowed for two tiles that occupy given positions in $\mathbb{R}^n$.  Matching rules are typically taken to be locally specifiable.  For present purposes, the matching rules may constrain the pairs of tile types allowed for adjacent tiles and for next-nearest-neighbor tiles. 
\item A \emph{tiling model} is an assignment of energies to the tilings that can be composed from a given set of tiles.  We construct tiling models in which the energies are determined by the number and type of violations of matching rules in the given tiling, with each violation independently contributing a positive definite energy.
\item A \emph{prototile} \cal{P} is a prototype of a tile.  It is a geometric unit that is shaped or decorated in a way that displays the matching rules directly.  Each tile in a tiling is a copy of a prototile.  Each different tile type can be realized as a rotations and/or reflection of one element of the set of the tiling's prototiles, $\{\cal{P}_i\}$, $i = 1, \ldots K$.
\item A \emph{lattice model} assigns a generalized spin variable, $q_j$, to each of the sites of a dicrete lattice, and an energy to each spin based on its value and those of the spins in its local environment.  For a tiling model in which the geometric arrangement of tiles is a lattice (though the tile types are not determined by the lattice structure), a lattice model can be constructed in which the spin index $q_j$ indicates the type of tile at lattice site $j$.  The Hamiltonian for such a lattice model assigns an energy to each spin that corresponds directly to the energy of the corresponding tile in the tiling model.
\end{itemize}

The term ``tiling model'' is chosen intentionally to suggest that the model could be realized physically by a collection of units whose shapes correspond to the tile shapes and whose internal structure imposes energetic biases that enforce the matching rules.  In the physical system, the assembly of tiles into the close-packed structures of interest is an important part of the assembly process, but one that we do not study in the present work.  Instead, we study the associated lattice models.  That is, we study the thermodynamic stability of the orientations of the tiles, given that they are already packed into the correct lattice structure but allowed to rotate in place and allowed to convert from one enantiomorph to the other.

\begin{figure*}
\begin{center}
\includegraphics[width=\textwidth]{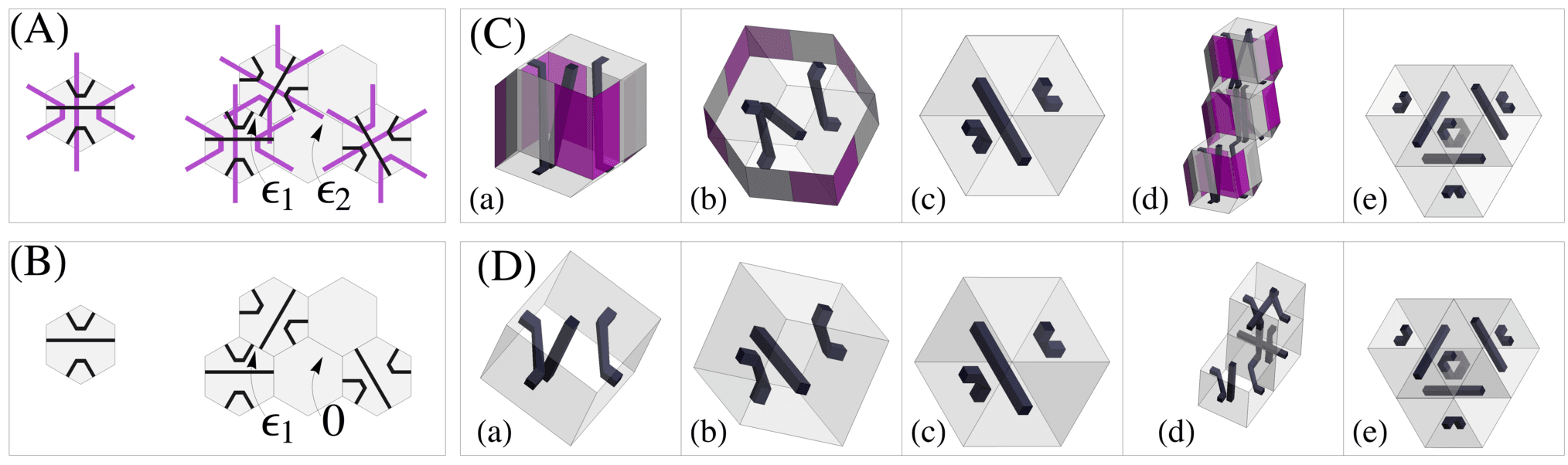}
\caption{(Color online.)
{\bf (A)} The Taylor-Socolar model prototile and matching rules.   Discontinuities in the black and thick gray (purple) stripes have energetic penalties $\e_1$ and $\e_2$, respectively.
{\bf (B)} The black stripe model prototile and matching rules.   Discontinuities in the black stripes have energetic penalties $\e_1$.  The model is equivalent to the Taylor-Socolar model with $\e_2 = 0$.
{\bf (C)} The zonohedral model.  (a), (b), (c) Different views of the prototile.  Each black bar reaches from one top face of the zonohedron to one of the bottom faces.  Panel (c) shows the view down the 111 axis.  (d), (e) Two views of three tiles in separate layers.  The black bars connecting through the shared faces of the tiles form a helix.  Viewed along the 111 axis, the helix is seen to correspond to a small triangle in the Taylor-Socolar tiling.
{\bf (D)}  The cubic model.  The views shown correspond to those in the bottom left panel. 
\label{fig:tilingmodels}}
\end{center}
\end{figure*}

We consider four distinct tiling models, each based on a tiling with a single prototile:

\noindent
{\bf Taylor-Socolar model:}  The prototile is a 2D hexagon with markings that break all of its rotation and reflection symmetries.  There are 12 tile types, corresponding to the 6 rotations and 2 reflections of the prototile.  The matching rules govern nearest-neighbor and next-nearest-neighbor tile pairs, and can be conveniently expressed using the decoration shown in Fig.~\ref{fig:tilingmodels}(A).  
The rule is that all black and purple line in the tiling must join to form continuous lines.   For obvious geometric reasons, the tilings are all close-packed hexagonal structures.  For each pair of adjacent tiles, a positive energy $\e_1$ is assigned if the black stripe is not continuous across the shared boundary, and for each pair of next-nearest-neighbor tiles, a positive energy $\e_2$ is assigned if the relevant purple stripe along the intervening tile edge is not continuous.  Viewed as a 2D model, the Taylor-Socolar prototile is chiral and both enantiomorphs are needed.  Viewed, however, as a 2D layer of a 3D system, the prototile is not chiral, as a rotation by $\pi$ about an in-plane axis converts one 2D enantiomorph into the other.  The corresponding lattice model is a triangular lattice with 12 possible values for $q_j$ and both nearest and next-nearest neighbor interactions.

\noindent
{\bf Black stripe model:} \label{sec:eps2eq0}  The prototile is the Taylor-Socolar prototile with the next-nearest-neighbor matching conditions removed.  The model is equivalent to the Taylor-Socolar model with $\e_2=0$. (See  Fig.~\ref{fig:tilingmodels}(B).)  In this model, the prototile is \emph{not} chiral; the tiling consists of a single tile type and its rotations.  Thus we have only 6 tile types.  The corresponding lattice model is a triangular lattice with 6 possible values for $q_j$ and only nearest neighbor interactions.

\noindent
{\bf Zonohedral model:}  The prototile is a rhombic dodecahedron with markings as shown in Fig.~\ref{fig:tilingmodels}(C).  The edges of the prototile lie along the tetrahedral directions $111$, $1\overline{1}\overline{1}$, $\overline{1}1\overline{1}$, and $\overline{1}\overline{1}1$.  The prototile is chiral, and the tiling contains both enantiomorphs.  This model is closely related to the Taylor-Socolar model.    The tiles sit at the sites of a face-centered cubic lattice.  The matching rules are that darkest gray (purple) and gray patches around the equator must match to like colors and black bars must be continuous across faces,  and the energetic costs for mismatches are $\e_1$ and $\e_2$, respectively.  Orienting the lattice such that the 111 direction is vertical, each layer of tiles at the same height forms a hexagonal packing in which the color matching rule is equivalent to the purple stripe rule in the Taylor-Socolar model.  The black bars connect tiles in different layers.  They are almost equivalent to the Taylor-Socolar black stripes, but there is a subtle difference.  Because the bars connect tiles in different layers, there cannot be a closed triangle.  As shown in panels (d) and (e) at bottom right in Fig.~\ref{fig:tilingmodels} and explained in detail in Section~\ref{sec:zonohedral}, a triangle in the Taylor-Socolar tiling becomes an infinite helix in the zonohedral tiling.  The corresponding lattice model is a face-centered cubic lattice with 12 possible values for $q_j$ and only nearest-neighbor interactions.  Note, however, that the prototiles can be compressed as desired along the 111 direction and the length of the 6 edges oriented along the 111 direction is arbitrary as well.

\noindent
{\bf Cubic model:}  The prototile is a rhombohedron with markings as shown in Fig.~\ref{fig:tilingmodels}(D).  The length of the diagonal in the 111 direction can be chosen arbitrarily; we take it to correspond to a cubic tile shape for convenience.  The prototile is chiral, and the tiling contains both enantiomorphs. This model is equivalent to the zonohedral tiling with the color matching rules deleted; i.e., with $\e_2=0$.  It is thus related to the black stripe model in the same way that the zonohedral model is related to the Taylor-Socolar model; the black bars in the tiling can form helices whose projections on the 111 direction are the triangles in a black stripe tiling.  The corresponding lattice model is a simple cubic lattice with 6 possible values for $q_j$ and only nearest-neighbor interactions.   

%
\section{2D models}\label{sec:2Dmodels}
The Hamiltonian for the Taylor-Socolar lattice model assigns an energy $\e_1>0$ to each nearest-neighbor pair of tiles sharing an edge where the black stripe matching rule is violated, and similarly, an energy $\e_2>0$ to pairs or next-nearest-neighbor tiles for which the purple stripe matching rule is violated.  Pairs for which the matching rules are satisfied (the stripe decorations are continuous) are assigned zero energy.  In the following, we set $\e_1$ as the unit of energy and temperature; that is, we take $\e_1=1$ and the Boltzmann constant $k_B=1$.

\subsection{Order parameters}\label{sec:orderParameters}
\subsubsection{Level 1}\label{sec:level1op}
\begin{figure}
\centering
\includegraphics[width=1.0\columnwidth]{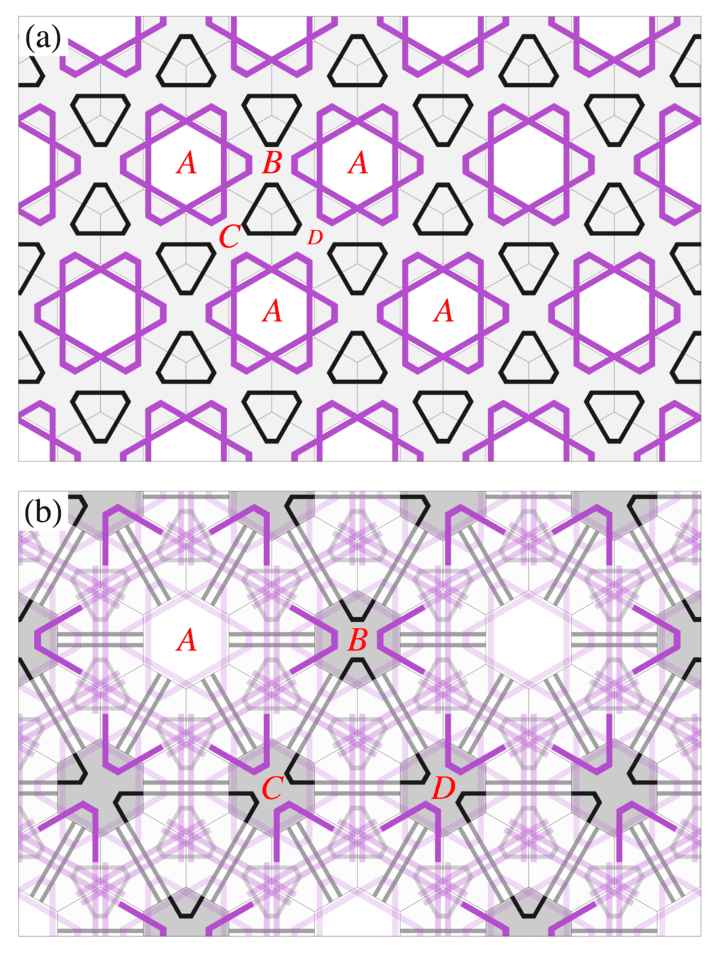}
\caption{(Color online.) (a) Level-1 ordering in the Taylor-Socolar model.  A subset comprising three quarters of the tiles is shown.  For each tile, the black and thick gray (purple) corner decorations are included, but not the long stripes.  Each tile shown may be in any of the four orientations corresponding to the possible positions of the long black and thick gray (purple)  stripes.  The tiles lying on sublattice $A$ do not contribute any of the decorations in the pattern shown here.  (b) Level-2 ordering in the Taylor-Socolar model.  Three quarters of the tiles on sublattice $A$ of panel (a) participate in the formation of black and purple triangles.  The light-colored tiles and decorations display the level-1 order.  Double stripes indicate the possible locations of black and thick gray (purple) stripes on tiles that contribute corners to the level-1 triangles.
\label{fig:levels1and2order}}
\end{figure}

The Taylor-Socolar lattice model undergoes an unusual second order phase transition at temperature $T_{c;1}\approx 1.51$.  Below $T_{c;1}$, three quarters of the tiles lock into orientations forming honeycomb lattices of small (truncated) triangles of both the black stripes and purple stripes, as shown in Fig.~\ref{fig:levels1and2order}(a).  Note that the purple stripes form three overlapping, scaled and rotated copies of the black stripe pattern.~\cite{SocolarTaylor2010a} 
The remaining quarter of the tiles, which occupy the sites of sublattice $A$, have no preferred orientation.  
We refer to the tiles that form the level-1 order as the ``corner set,'' and the tiles that have no preferred orientation just below the transition temperature as the ``rattlers.''

\begin{figure}
\begin{center}
\includegraphics[width=0.8\columnwidth]{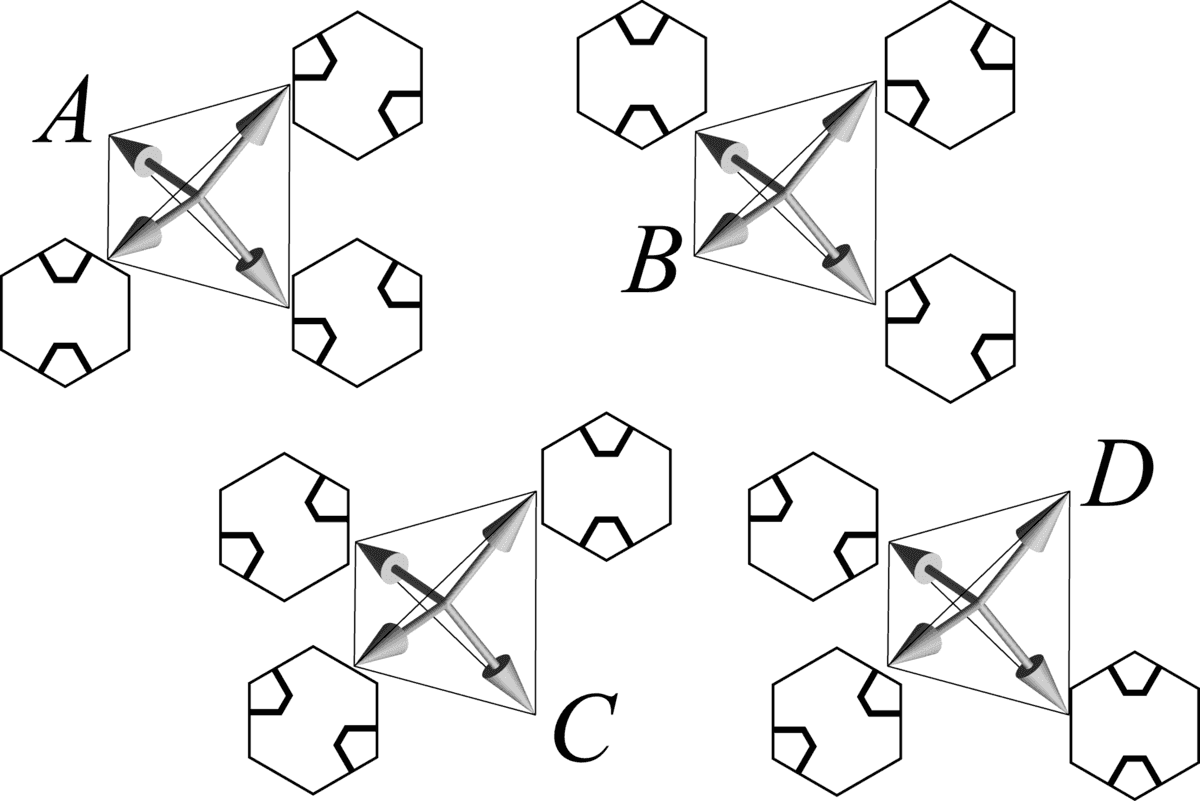}
\caption{The spins used to define the order parameter for the level-1 transition.  See text for explanation.
\label{fig:op}}
\end{center}
\end{figure}

An order parameter for the transition was defined in Ref.~\cite{Byington2012}.  Each tile $j$ is assigned a ``staggered tetrahedral spin'' vector ${\bm \sigma}_{1,j}=\ev_X$, where $X$ indicates one of the four vertices of a reference tetrahedron (see Fig.~\ref{fig:op}.)
The spin is determined both by the orientation of the diameter joining its two black triangle corners and by the sublattice to which it belongs, according to the map shown in Fig.~\ref{fig:op}.   For example, a tile with corners aligned vertically and sitting on the $B$ sublattice is assigned $\sigma_{1}=\ev_A$.  Note that specifying ${\bm \sigma}_{1,j}$ does not completely specify the orientation of tile $j$.  There are four consistent choices, corresponding to the two possible locations of the long black stripe and two possible orientations of the long purple stripe.   Note also that for any given tile, ${\bm \sigma}_{1}$ can take only three of the four possible values.

We define the average total spin ${\bm \sigma}_{1,{\rm tot}} \equiv \frac{1}{N}\sum_{j} {\bm \sigma}_{1,j}$, where $N$ is the number of tiles in the system.  In the pattern shown in Fig.~\ref{fig:levels1and2order}(a), which only consists of the $B$, $C$, and $D$ sublattices, the total spin lies in the ${\bm e}_A$ direction.  Alternatively, the pattern could form around the sites of the $B$, $C$, or $D$ sublattice, yielding ${\bm \sigma}_{1,{\rm tot}}$ in the corresponding direction.

The system exhibits tetrahedral symmetry in the following sense:  for each configuration with a given ${\bm \sigma}_{1,{\rm tot}}$, there is another with identical energy having ${\bm \sigma}_{1,{\rm tot}}^{\prime}$ related to ${\bm \sigma}_{1,{\rm tot}}$ by an operation in the 24-element tetrahedral group $T_d$.  The mapping from operations on the lattice to elements of $T_d$ is given in Table~\ref{tab:group}.

\begin{table}[t]
\begin{center}
\begin{tabular}{|ccc|}
\hline
\parbox[t]{1.4in}{\raggedright Lattice operation} & \  & \parbox[t]{1.1in}{\raggedright $T_d$ operation on ${\bm \sigma}$} \\ [4pt]
\hline
\parbox[t]{1.4in}{\raggedright Rotation by $2\pi/3$ \\ $\quad$ about center of $X$} & $\rightarrow$ & \parbox[t]{1.1in}{\raggedright Rotation by $2\pi/3$ \\ $\quad$about ${\bm e}_X$} \\ [12pt] \hline
\parbox[t]{1.4in}{\raggedright Reflection through edge \\ $\quad$ shared by $X$ and $Y$} & $\rightarrow$ & \parbox[t]{1.1in}{\raggedright Reflection through \\ $\quad$ $({\bm e}_X,{\bm e}_Y)$ plane} \\  [12pt] \hline
\parbox[t]{1.4in}{\raggedright Translations taking \\ $\quad$ $X$ sublattice to $Y$ }& $\rightarrow$ & \parbox[t]{1.1in}{\raggedright Rotations by $\pi$ \\ $\quad$ about ${\bm e}_X+{\bm e}_Y$} \\ [12pt] \hline
\parbox[t]{1.4in}{\raggedright Rotation by $2\pi/3$ \\ $\quad$ followed by reflection} & $\rightarrow$ & \parbox[t]{1.1in}{\raggedright Rotary inversion} \\ [12pt]
\hline
\end{tabular}
\caption{Symmetry operations for the total staggered tetrahedral spin.  The left column specifies an operation on the 2D tiling pattern, where $X,Y \in  \left\{ A, B, C, D\right\}$ each represent a tile in the corresponding sublattice of Fig.~\ref{fig:op}.  The right column specifies 3D operations on the order parameter in terms of  the tetrahedral star of vectors ${\bm e}_X$, where $X$ is the label shown on Fig.~\ref{fig:op}.}
\label{tab:group}
\end{center}
\end{table}

The order parameter for the transition is
\begin{equation}
\phi_1 = \max\left({\bm \sigma}_{1,{\rm tot}}\cdot\ev_X\right)\,,
\end{equation}
where $X$ runs over the sublattice indices $\{A, B, C, D\}$.
The projection operation in the definition of $\phi_1$ serves to assign the same value to all configurations with the same tile orientations in the corner set, but different rattler configurations.
 
\subsubsection{Higher levels}\label{sec:levelnop}
When level 1 is fully ordered, the tiles on one of the four sublattices remain free to rotate.  For the example shown in Fig.~\ref{fig:levels1and2order}(a), these are the white tiles on sublattice $A$.  One sees by inspection, however, that those tiles form a lattice equivalent to the original lattice and with an equivalent matching rule enforced through the long stripes on the level-1 tiles that connect the tiles of sublattice $A$, as illustrated in Fig.~\ref{fig:levels1and2order}(b).  We therefore define a second order parameter, $\phi_2$, analogous to $\phi_1$ but obtained by summing only over the tiles in sublattice $A$ (which is now regarded as a union of four sparser lattices).  Given a full ordering of $\phi_2$, one can then identify the correct sublattice for defining $\phi_3$, and so forth.  Each order parameter $\phi_n$ thus measures the degree to which a periodic lattice of black triangles with edges consisting of $2^{n-1}-1$ tiles is formed.
 
\subsection{Monte Carlo results for slow quenches}
We implement a Monte Carlo simulation of the Metropolis algorithm, involving only moves that change the orientation of a single tile, to study the phase transitions.~\cite{metropolis-etal_JChemPhys53,NewmanBarkema}  Throughout this paper, we define one Monte Carlo step (MCS) to be $N$ attempted Metropolis moves, where $N$ is the number of tiles in the system.  The system is taken to be a rhombus with $\ell$ tiles per edge and periodic boundary conditions.  Each tile $i$ is assigned an energy 
\begin{equation}
U_i(o) = \e_1 m_b + \e_2 m_p \,,
\end{equation} 
where $o$ indicates the orientation of the tile and $m_b$ and $m_p$ are the numbers of mismatches in the black and purple stripes, respectively, among the tile pairs that include tile $i$.

\begin{figure*}[t]
\begin{center}
\includegraphics[width=1.0\textwidth]{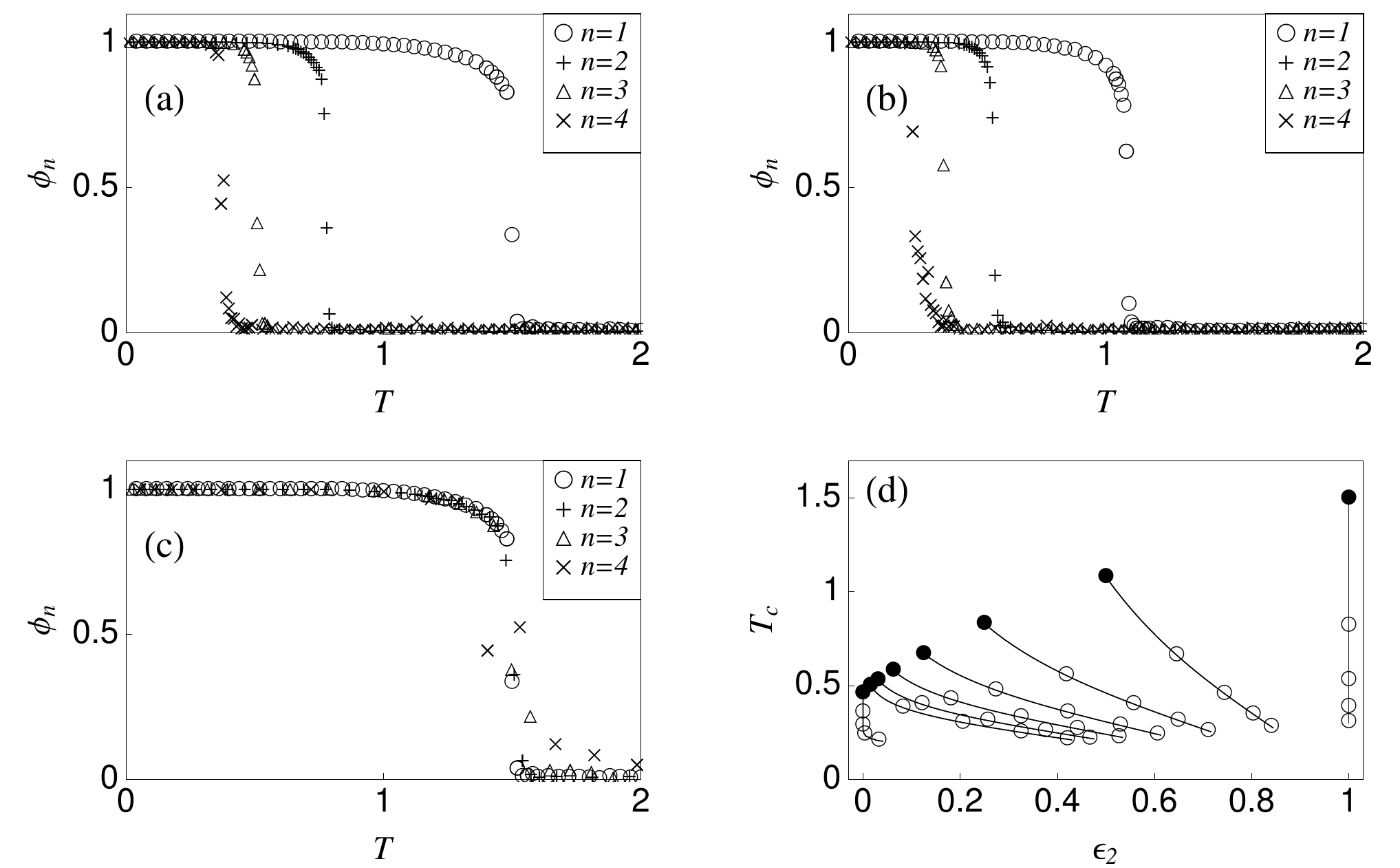}
\caption{
(a) The order parameters $\phi_n$ vs $T$ from a quench for the case $\e_2=1$, with quench parameters $\ell=64$, $T_0=2.0$, $T_f=0.0$, $\Delta T=0.01$ and $\tau=12\times 10^5$ MCS. 
(b) The order parameters $\phi_n$ vs $T$ from a quench for the case $\e_2=0.5$, with the same quench parameters as in (a).
(c) Data collapse obtained from the scaling theory for the data from (a).  Deviations of the level-4 points from the others are finite-size effects due to the relatively small number of level-4 triangles in the system. 
(d) Solid circles show the dependence of $T_{c;1}$ on $\e_2$, for $\e_2\leq 1$, obtained from simulations of slow quenches. For each value of $\e_2$, $T_{c;1}$ is approximated as the highest temperature for which $\phi_1>0.1$.  Open circles show the dependence of $T_{c;n}$ on $\e_2$, for $\e_2\leq\e_1=1$, obtained from Eq.~\eqref{eq:scaling}.  The lines in this figure connect parameters for systems that exhibit equivalent behavior.  As one follows a line downward and to the right, the open circles indicate transitions of levels $n=2$ through $n=5$. 
\label{fig:opvTandScaling}}
\end{center}
\end{figure*}

A tile $i$ is selected at random.  Let $q$ denote its current orientation.  A proposed new orientation, $q^{\prime}$, is selected at random from its 12 possible states.  The transition is accepted with probability 
\begin{equation} \label{eq:probability} 
P(i,q,q^{\prime}) = \min\la\{1,\exp\la(-(U_i(q^{\prime})-U_i(q))/T\ra)\ra\} \,.
\end{equation} 
A slow quench from some $T_0$ to some $T_f$ is simulated by reducing $T$ in steps of $\Delta T$, sitting for a time $\tau = 12\times 10^5$ MCS at each step.  The results do not change if we use longer equilibration times at each step.  For each $T$, we compute the average values of the order parameter and of the energy density.  Figure~\ref{fig:opvTandScaling}(a) shows strong evidence of a second order phase transition.  Consistent with this expectation, our Monte Carlo simulations of slow quenching and reheating show no evidence of hysteresis.  Based on the tetrahedral symmetry of the system, we expect the transition to be in the universality class of the 4-state Potts model.  This universality class has  a very small order parameter exponent $\beta = 1/12$, which appears to be consistent with the data, although it is exceedingly difficult to obtain a clean numerical determination of such a small value of $\beta$.  

Fig.~\ref{fig:opvTandScaling}(a) shows the values of the order parameters $\phi_n$ at the sampled temperatures.  The data shown here are from a simulation with $\e_2=1$, $\ell=64$, $T_o=2.0$, $T_f=0.0$, and $\Delta T=-0.01$.  An important feature of the plot is the rapid saturation of $\phi_1$ below $T_{c;1}$.
Fig.~\ref{fig:opvTandScaling}(b) shows the same plot of $\phi_n$ from a simulation with $\e_2=0.5$. As might be expected, the phase transitions in the hierarchy occur at lower values of $T_{c;n}$, and in general the transition temperatures are lower for smaller values of $\e_2$, as indicated by solid circles in Fig.~\ref{fig:opvTandScaling}(d).  Remarkably, however, the transition temperatures  do not go to zero for $\e_2=0$, nor are these transitions preempted by a transition to a different phase; the sequence of transitions leading to the limit-periodic state still takes place despite the existence of periodic ground states for this Hamiltonian.

\subsection{Scaling relations for the transition hierarchy}
\label{sec:scalingRelations}
To better understand these hierarchy of transitions, we explore the partition function of the separate levels of the limit periodic system.
The partition function of the entire system can be written as a configuration sum of the following form:
\begin{equation}
Z_1(T;\e_1,\e_2) = \sum_{\rm config.}\prod_{\rm n.n.}e^{(\e_1\pm\e_1)/2T}\cdot\prod_{\rm n.n.n.}e^{(\e_2\pm\e_2)/2T}.
\end{equation}
Here the products are over nearest-neighbor and next-nearest-neighbor bonds, respectively, and the sign in the exponent is taken to be positive if that bond is mismatched in the current configuration and negative if that bond is matched.

Now, let us assume for the moment that $\phi_1$ is fully saturated; i.e., that the tiles forming the level-1 lattice are somehow clamped into the configuration of triangle corners shown in Fig.~\ref{fig:levels1and2order}(a).  The remaining triangle corners that are free to move now lie on the tiles of sublattice $A$.  These corners do not connect directly, but do become correlated due to an effective interaction mediated by the long black stripes on the tiles of the $B$, $C$, and $D$ sublattices.  In fact, the partition function for the remaining degrees of freedom in the tiling at a given temperature $T_1$ is precisely equivalent to the original partition function, but with renormalized values of $\e_1$, $\e_2$, and $T$.

Under the saturation assumption, these bonds are independent in a given configuration of the level $n$ tiles.  Hence, the partition function of the entire configuration can be written as a product of the appropriate $\zeta_\pm$.  Using the same configuration sum as in the full level-1 case, the level-$n$ partition function is written in the form:
\begin{equation}
Z_{n}(T;\e_1,\e_2) = \sum_{\rm config.}\prod_{\rm n.n.}\zeta_n^{\pm}(T;\e_1)\cdot\prod_{\rm n.n.n.}\zeta_n^{\pm}(T;\e_2),
\end{equation}
Again the value of each $\pm$ is determined by the state of the bond, matched or mismatched, in the configuration being summed.

Due to the identical configuration sums in the partition functions, the level-$n$ system behaves equivalently to the system at level-1 when the level-$n$ bond partition functions $\zeta_n^{\pm}(T;\e)$ are equal to those for level-1 $\zeta_1^{\pm}(T;\e)\equiv e^{-\e/T}$, up to a constant scaling factor.  We exploit this relation to determine the scaling factors for $\e$ and $T$.

Explicitly, the partition function for level $n$ is identical to that of an effective level-1 system if and only if the following system of equations holds:
\begin{eqnarray}
\zeta_n^{+}(T_n;\e_1)&=\alpha_1 \zeta_1^{+}(T_1;\e_1) \nonumber\\
\zeta_n^{-}(T_n;\e_1)&=\alpha_1 \zeta_1^{-}(T_1;\e_1) \nonumber\\
\zeta_n^{+}(T_n;\e_{2;n})&=\alpha_2 \zeta_1^{+}(T_1;\e_{2;1}) \label{eq:scsys}\\
\zeta_n^{-}(T_n;\e_{2;n})&=\alpha_2 \zeta_1^{-}(T_1;\e_{2;1}) \nonumber.
\end{eqnarray}

\begin{figure}[t]
\centering
\includegraphics[width=.7\columnwidth]{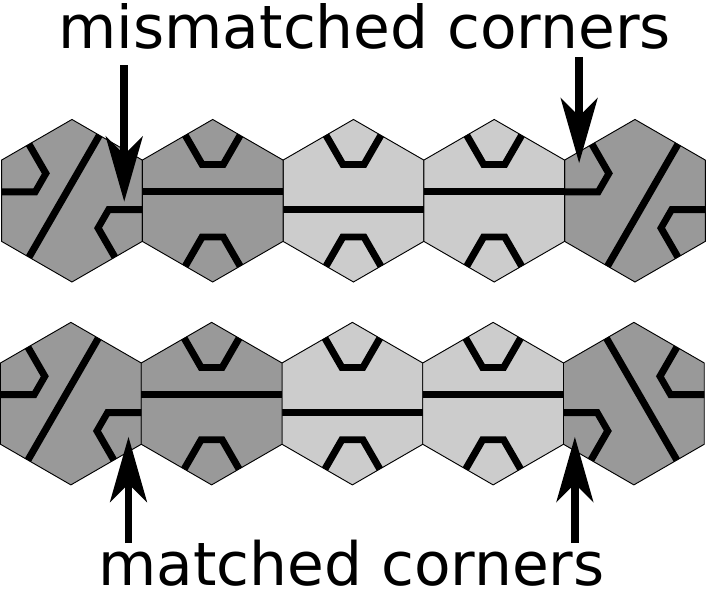}
\caption{ Matched and mismatched corner configurations for level-3 edges. 
\label{fig:level3bond}}
\end{figure}

Here $\alpha_1$ and $\alpha_2$ are arbitrary constants and we assume that $\e_1$ is fixed for the scaling (it serves as our unit of energy).  To reduce Eq.~\eqref{eq:scsys} to a scaling relation for $\e_2$ and $T$, we note that each level-$n$ bond is a 1D Ising chain with $2^{n-1}$ possible mismatches, as illustrated in Fig.~\ref{fig:level3bond}.  Therefore the level-$n$ bond partition functions are simply
\begin{eqnarray}
\zeta_n^{-}(T;\e) & =  &\frac{1}{2}\left[\left(1+e^{-\e/T}\right)^{k_n}\!\! - \left(1-e^{-\e/T}\right)^{k_n}\right]; \nonumber \\
\zeta_n^{+}(T;\e) & = &\frac{1}{2}\left[\left(1+e^{-\e/T}\right)^{k_n} \!\! + \left(1-e^{-\e/T}\right)^{k_n}\right], 
\label{eq:nbpart}
\end{eqnarray}
where $k_n\equiv2^{n-1}$.  Equations.~\eqref{eq:scsys} and~\eqref{eq:nbpart} imply the following scaling relations for $T_n$ and $\e_{2;n}$:
\begin{eqnarray}
\label{eq:scaling}
\tanh\left(\frac{\e_1}{2T_1}\right) & = & \left[\tanh\left(\frac{\e_1}{2T_n}\right)\right]^{k_n} \nonumber \\ 
{\rm and} \quad
\tanh\left(\frac{\e_{2;1}}{2T_1}\right) & = & \left[\tanh\left(\frac{\e_{2;n}}{2T_n}\right)\right]^{k_n}.
\end{eqnarray}

The scaling relations apply for all $T$.  Consider now the behavior of the system during a slow quench.  When $T$ drops below $T_{c;1}$, the level-1 ordering rapidly sets in.  For the case $\e_2=\e_1$, Eq.~\eqref{eq:scaling} immediately implies $\e_{2,n}=\e_{2;1}$.  Thus in this case, Eq.~\eqref{eq:scaling} gives a relation between the behaviors of the same system at different temperatures.  Recall that this relation is derived under the assumption that the level-($n-1$) order is perfectly locked in at all temperatures for which $\phi_n$ is nonzero.  For the renormalized temperature 
\begin{equation}
T_{c;2} =  2 \left[ \tanh^{-1}\left(\sqrt{\tanh(1/2T_{c;1})}\right)\right]^{-1},
\end{equation} 
at which the level-2 partition function maps onto the level-1 partition function at $T_{c;1}$, we find $\phi_1(T_{c;2})=0.992$, so the deviations from the derived relation are expected to be small.  A more detailed study of these deviations is discussed (in the context of a different model) in Section~\ref{sec:freeenergyscaling}.   

Using the scaling relations in Eq.~\eqref{eq:scaling} for the case $\e_2 = \e_1$, we obtain an excellent data collapse for several levels by plotting $\phi_n(T_n)$ as a function of $T_1(T_n)$, as is seen in Fig.~\ref{fig:opvTandScaling}(c).
These scaling relations also yield predictions when $\e_2\neq\e_1$.  In this case, holding $\e_1$ fixed, one can map the level-$n$ system at a given $\e_2$ and $T_n$ onto the level-$1$ system at a different $\e_2$ and $T_1$ by solving the first equation for $T_1$ and the second for $\e_{2;1}$.  The structure of the scaling relations is shown in Fig.~\ref{fig:opvTandScaling}(d).  Each circle in the figure marks a critical temperature for some transition.  Points connected by a line are equivalent by the scaling relations, with the level increasing as one moves down and to the right.  The curves all approach the point $T=0$ and $\e_2=1$ as $n\rightarrow \infty$.

\subsection{The black stripe model: $\e_2 = 0$}\label{sec:blackstripe}
\begin{figure}
\begin{center}
\includegraphics[width=1.0\columnwidth]{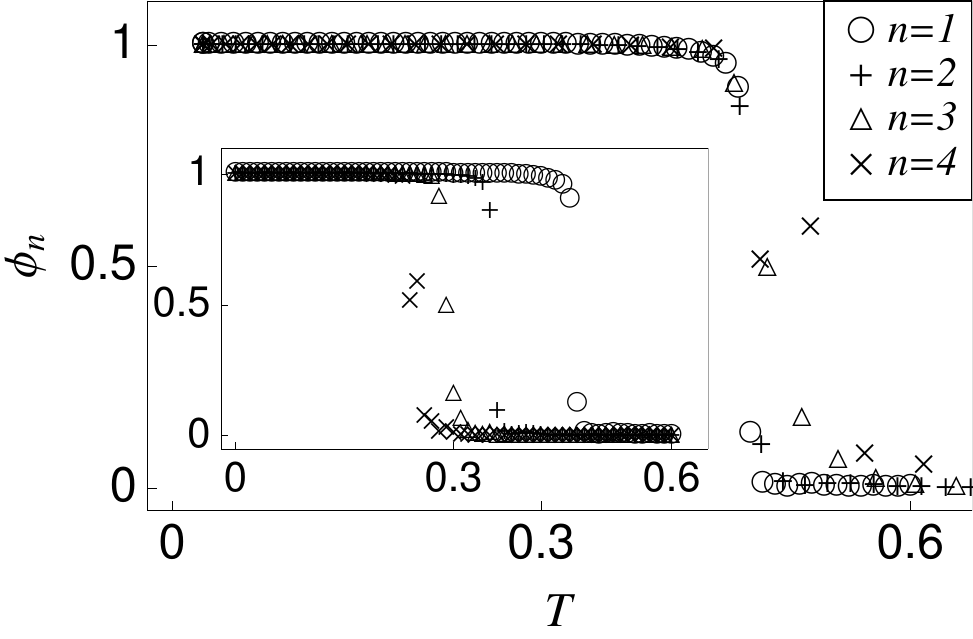}
\caption{Inset:  $\phi_n$ vs $T$ for the black stripe model, $\e_2=0$. The data is from simulated quenching on a rhombic domain of side length 64, with parameters $T_o=0.6$, $T_f=0.0$, $\Delta T=0.01$ and $\tau=12\times 10^5$ MCS.  Full panel:  Data collapse from the scaling theory for the black stripe model applied to the data from the inset.
\label{fig:opvTzero}}
\end{center}
\end{figure}

As illustrated in Fig.~\ref{fig:opvTandScaling}(d), the scaling relations of Eq.~\eqref{eq:scaling} imply that for any $\e_2 <  \e_1$, the effective value of $\e_2/\e_1$ approaches $0$ in the limit of large $n$.  It is thus important to study the transition more carefully for the $\e_2 = 0$ case.  Figure~\ref{fig:opvTzero} strongly suggests that the level-1 transition does occur.  Furthermore, for $\e_2 = 0$, the scaling relations for $n>1$ reduce to the same simple form as for the $\e_2 = \e_1$ case, consistent with the collapse shown in  Fig.~\ref{fig:opvTzero}.  At the critical temperature for the level-2 transition ($T^*_{c;2}\approx 0.365)$, we have $\phi_1\approx 0.998$. 

In Section~\ref{sec:cubic}, we study a 3D analogue of the black stripe ($\e_2=0$) model in much greater detail.

%
\section{3D models}\label{sec:3Dmodels}
One motivation for considering 3D models is that the next-nearest-neighbor interactions between the 2D tiles can be realized in a natural way as nearest-neighbor interactions in 3D tiles.  There is, however, an important difference between the zonohedral model and the Taylor-Socolar model, as mentioned in Section~\ref{sec:defns}.  A feature of the Taylor-Socolar model that played a significant role in the proofs of aperiodicity of the ground state~\cite{SocolarTaylor2010a} is that when two tiles sharing a vertex are oriented such that two corners of a black triangle are formed around that vertex, the third tile sharing that vertex is forced to contribute a corner that completes the triangle.  The same is true for three next-nearest neighbor tiles that combine to form a thick gray (purple) triangle.  In the 3D models, however, the situation is not quite equivalent.

Let the tile centers in the 3D model be at the positions $j_1 \av_1 + j_2 \av_2 + j_3 \av_3$, where $j_i$ is an integer.  For the zonohedral model, which forms a FCC lattice, we take 
\begin{equation} \label{eqn:avFCC}
\av_1 = (1,1,0), \quad \av_2 = (0,1,1), \quad \av_3 = (1,0,1)\,.
\end{equation}
For the cubic model, which forms a simple cubic lattice, we take 
\begin{equation} \label{eqn:avSC}
\av_1 = (1,0,0), \quad \av_2 = (0,1,0), \quad \av_3 = (0,0,1)\,.
\end{equation}
Let $L_{\ell}$ denote the set of tiles in the layer defined by $j_1+j_2+j_3 = \ell$.  The black bar matching rules connect tiles in layer $\ell$ to tiles in layers $\ell+1$ and $\ell-1$, and never connect two tiles in the same layer, which immediately implies that the black bars cannot form triangles.  For a given black bar corner on a tile in layer $\ell$, the two tiles that contribute black bar corners connecting to it are not neighbors of each other, as one is in layer $\ell+1$ and the other is in layer $\ell-1$, and therefore do not constrain each others' orientations.

We show in the Appendix that this weakening of the matching rules still does not allow the set of ground states of the zonohedral model to include periodic tilings.  It is also straightforward to see that ground state configurations can be constructed that project directly onto the Taylor-Socolar model ground states, with the triangles in the 2D model becoming helices with axes along the 111 direction in the 3D models, although the proof does not yield a complete characterization of all of the degenerate ground states.  The Monte Carlo studies below indicate, however, that the ground state reached through slow quenching is in fact closely related to the Taylor-Socolar ground states.

\begin{figure}
        \centering
        \includegraphics[width = 1.0\columnwidth]{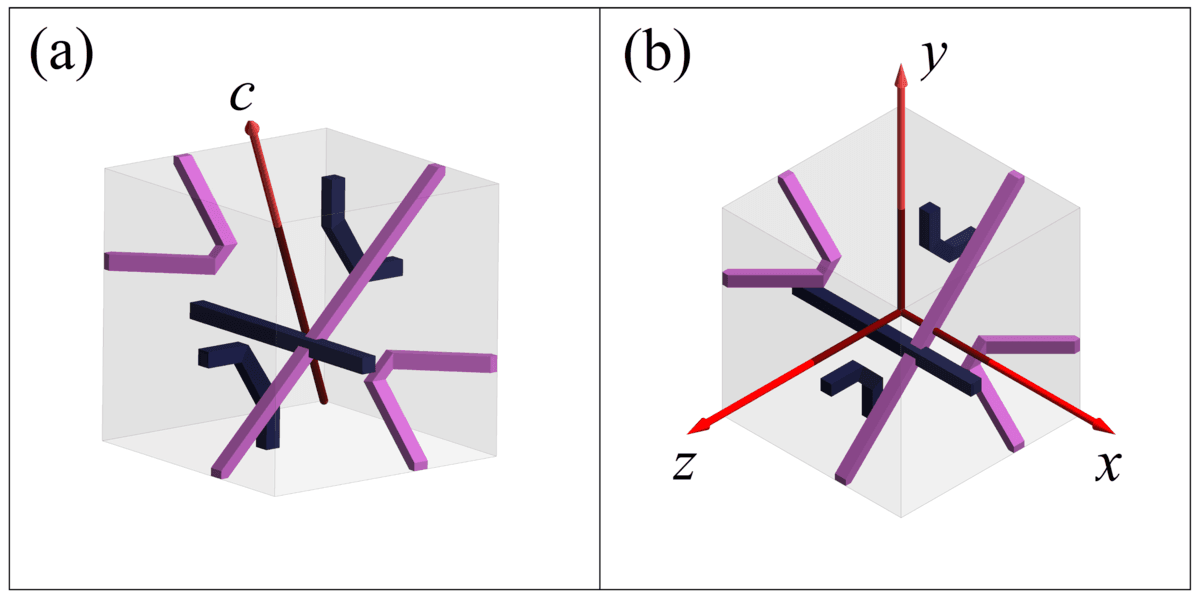}
        \caption{(Color online.)  Image of one enantiomorph of the cubic prototile, which is an alternate representation of the zonohedral tile. The axis arrows turn from dark to light where they intersect the faces of the cube.   (a.) The arrow indicates the 111 axis of the cube, defined to be the $c$ axis. (b.) A projection of the tile onto the plane perpendicular to the $c$ axis.
        \label{fig:cube}}
\end{figure}

For purposes of explication and visualization, we use an alternate version of the zonohedral tile in the discussion below.  We use both enantiomorphs of the chiral cubic prototile shown in Fig.~\ref{fig:cube}, which sit on the sites of the simple cubic lattice of Eq.~\eqref{eqn:avSC}.  In this representation, the colored faces of the zonohedral tile have been shrunk to zero height, so the color matching rules now appear as rules governing tiles that share an edge.  The rule is that the gray (purple) bars must continue across each edge.  As in the 2D model, matches are assigned an energy 0, black bar mismatches an energy $\e_1>0$, and purple bar mismatches an energy $\e_2>0$.

%
\subsection{The zonohedral model: $\e_2 = \e_1$}
\label{sec:zonohedral}

\begin{figure*}[tb]
\centering
\includegraphics[width = \textwidth]{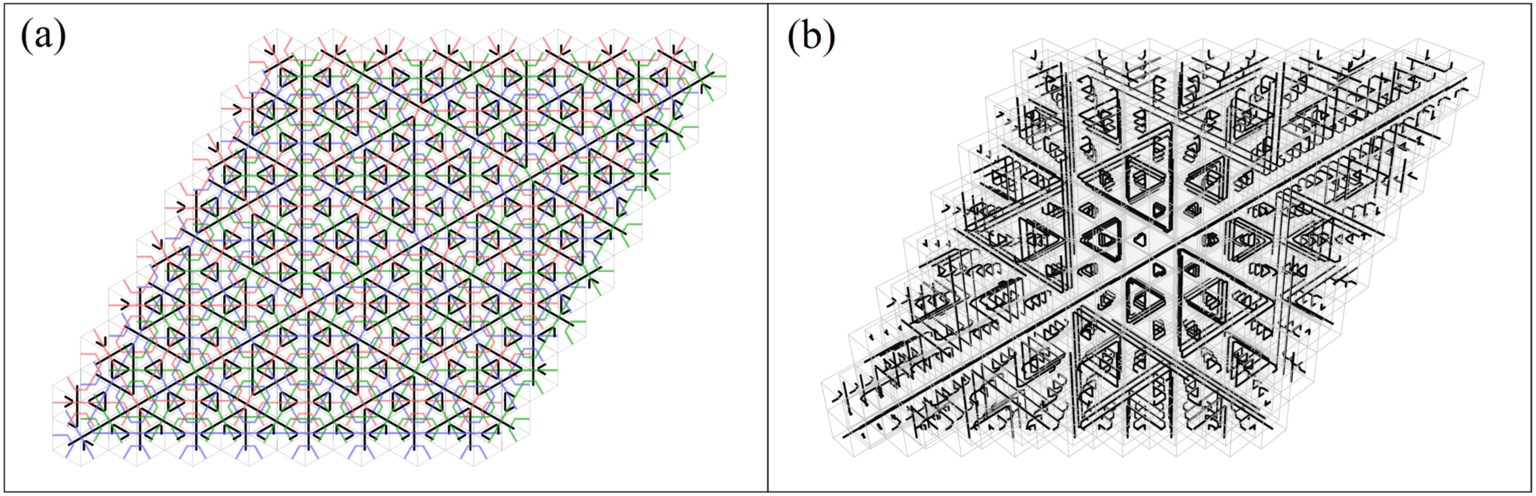}
\caption{(Color online.)  (a) A projection of a section of one of the limit-periodic states onto the plane perpendicular to the $c$-axis of the cubic lattice. The gray (red, green, and blue) triangles are formed from purple bars that were re-colored according to their layer for ease of viewing. Level-1 helices correspond to the smallest black triangles, level-2 helices to the next largest, and so on. Similarly, level-1 triangles are the smallest colored triangles, level-2 the next largest and so on.  (b) Image of 9 layers of the 3D structure that gives the projection of the black bars in (a). The gray (colored) bars were left out for clarity.  
\label{fig:2D}}
\end{figure*}
\subsubsection{Ground state structures}
\label{sec:groundState}

\begin{figure}
\centering
\includegraphics[width = 0.5\columnwidth]{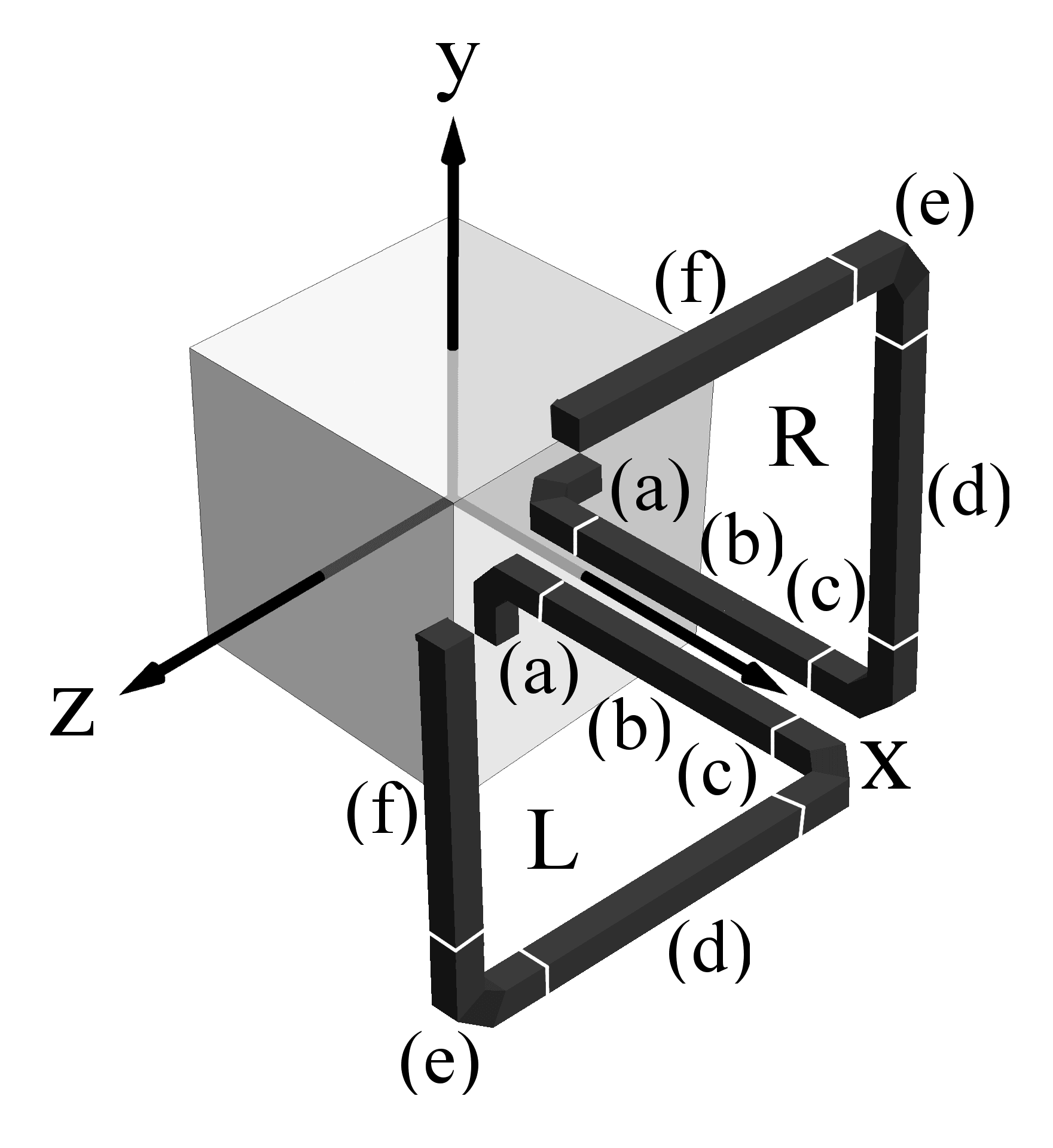}
\caption{Illustration of the structure of level-2 helices. Letters (a)-(f) label the parts of the helix, with (a), (c), and (e) being corners and (b), (d), and (f) being edges. The symbols R and L labels right- and left-handed helices, respectively.
       \label{fig:helix_figure}}
\end{figure}

In the zonohedral model, the ground states consist of parallel layers of tiles containing 2D patterns of purple bars identical to one of the three subsets of purple stripes in the Taylor-Socolar model.  These layers are coupled by black bars, which form arrays of helices aligned along the 111 axis (the $c$-axis of Fig.~\ref{fig:cube}(a)) whose projections onto the plane are triangles.  In one of the ground states, the projection of all of the  the black bars onto a plane normal to the $c$-axis is identical to the ground state of the Taylor-Socolar 2D model.  Images of this state are shown in Fig.~\ref{fig:2D}.
As in the 2D model, the triangles formed by the thick gray (purple) bars are labeled by an index $n$, such that a triangle of level-$n$ is formed by $3\cdot 2^{n-1}$ cubes. Similarly, we can group the black helices into levels such that a level-$n$ helix consists of $3\cdot 2^{n-1}$ cubes per turn.   Table~\ref{tab:helix} contains a complete description of a level-$n$ helix, and Fig.~\ref{fig:helix_figure} illustrates the level-2 case.
\renewcommand{\arraystretch}{1.5}
\begin{table}[tb]
\centering
\begin{tabular}{|c |c c|}
       \hline
        & coordinates of tiles & \\
       \hline
       \hline
      (a) & $(n_1,n_2,n_3)$ & $n_1, n_2, n_3 \in \mathbb Z$\\
       \hline
       (b) & $\{(n_1+x,n_2,n_3)\}$ & $1 \le x \le k^{\prime}_n,\,x\in\mathbb Z$\\
       \hline
       (c) &  $(n_1+k_n,n_2,n_3)$ & \\
       \hline
      (d) & $\mathrm{R:~} \{(n_1+k_n,n_2+y,n_3)\}$ & $1 \le y \le k^{\prime}_n,\,y\in\mathbb Z$\\
               & $\mathrm{L:~} \{(n_1+k_n,n_2,n_3+z)\}$ & $ 1 \le z \le k^{\prime}_n,\,z\in\mathbb Z$\\
       \hline
       (e) & $\mathrm{R:~} \{(n_1+k_n,n_2+k_n,n_3)\}$&\\
               & $\mathrm{L:~}  \{(n_1+k_n,n_2,n_3+k_n)\}$ &\\
       \hline
       (f) & $\mathrm{R:~}  \{(n_1+k_n,n_2+k_n,n_3+z)\}$ & $1 \le z \le k^{\prime}_n$\\
        &  $\mathrm{L:~} \{(n_1+k_n,n_2+y,n_3+k_n)\}$&$1 \le y \le k^{\prime}_n$\\
        \hline
\end{tabular}
\caption{Description of one turn of a single-stranded level-$n$ helix, where $k_n \equiv 2^{n-1}$ and $k^{\prime}_n\equiv 2^{n-1}-1$. Letters (a)-(f) label the parts of the helix, with (a), (c), and (e) being corners and (b), (d), and (f) being edges.  The coordinates listed locate the centers of the tiles that form these elements.  The label R (or L) indicates the set of coordinates for a right- (or left-) handed helix. Figure~\ref{fig:helix_figure} shows one full turn of a single strand of a level-2 helix for both the right- and left-handed cases.}
\label{tab:helix}
\end{table}

\begin{figure}
\centering
\includegraphics[width = 1.0\columnwidth]{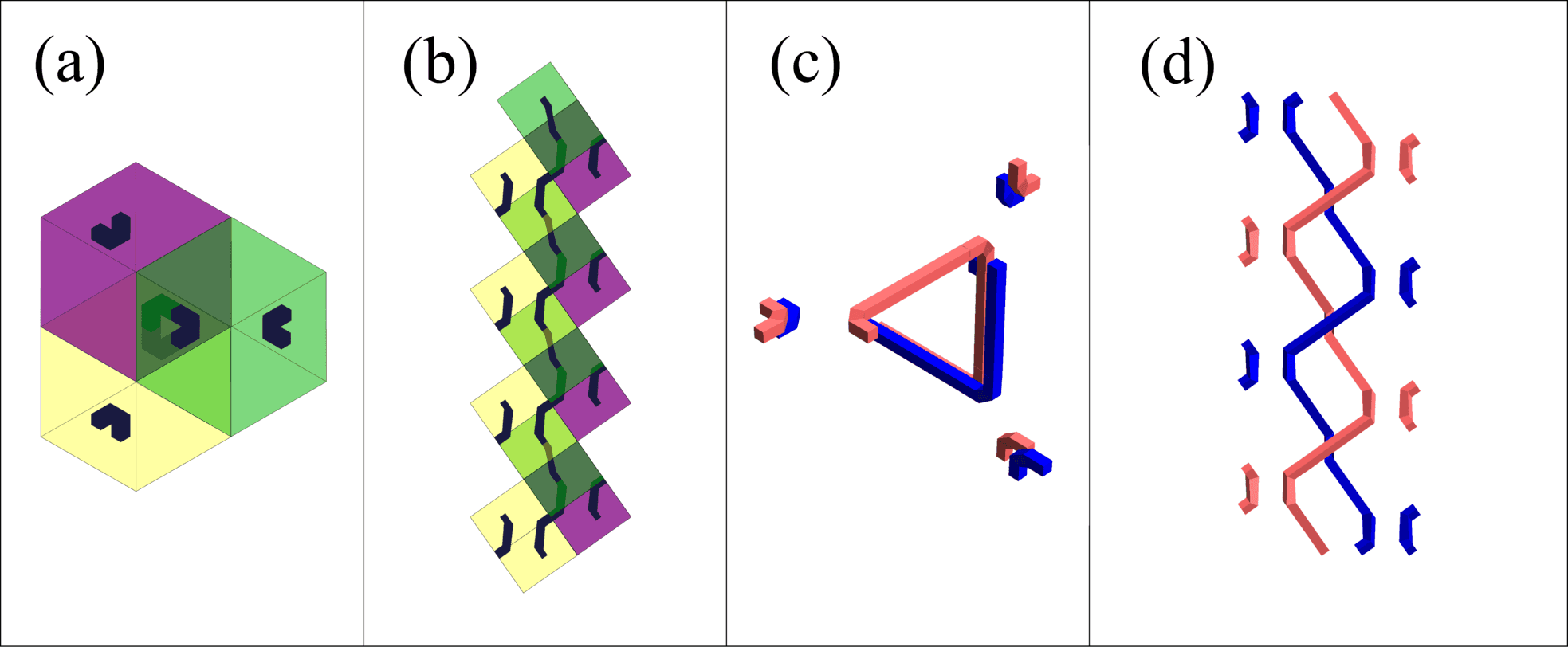}
\caption{(Color online.)  (a), (b) Representations of a level-1 helix. Cubes of the same color belong to separate columns.  (c), (d) Representation of a level-2 helix with level-1 designs left out for clarity.  Bars belonging to a given cube are shown in the same color.  Cubes containing bars of different colors do not interact with each other.
        \label{fig:column}}
\end{figure}

An ordered level-$n$ state is defined as a state in which helices at all levels with indices less than $n$ have ordered. We define a column of tiles to be the set of tiles at positions:
\begin{equation}
        \{(n_1+\ell, n_2+\ell, n_3+\ell)\}\mathrm{~for~} \ell \in \mathbb{Z}\,.
        \label{eq:kcol}
\end{equation}
Three columns forming a level-1 helix are depicted in Figs.~\ref{fig:column}(a) and~(b), with each column shown in a different color.

We define a subset of layer indices: 
\begin{equation}
p_n^{(i)} = \{p : \mathrm{mod}_{k_n} p = i \}\,,
\end{equation}
where $k_n = 2^{n-1}$.
The subset of layers with indices $p_n^{(i)}$ is denoted by $\ell_n^i$.  
Each level-$n$ helix has corners in layers $\ell \in \ell_n^i$ for some $i$.  The full set of level-$n$ helices is a union of $2^{n-1}$ lattices of helices, each corresponding to a different value of $i$.  Figures~\ref{fig:column}(c) and~(d) show a possible arrangement of two level-2 helices corresponding to the two different $i$'s.

\begin{figure}[tb]
       \centering
       \includegraphics[width =1.0\columnwidth]{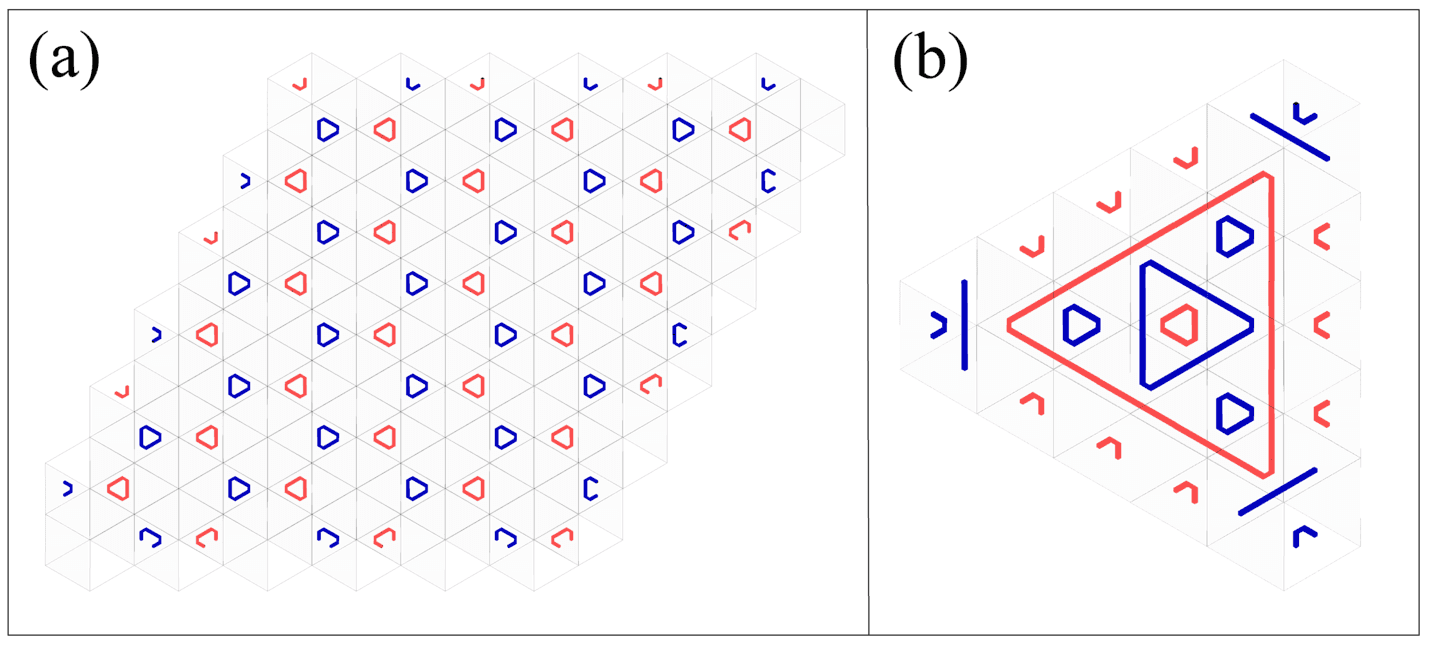}
       \caption{(Color online.) Helices of one chirality are colored gray (red), while the others are colored black (blue). (a) Level-1 lattice. (b) A section of a fully ordered structure.
       \label{fig:chirality_surroundings}}
\end{figure}

Consider now the level-$n$ helices corresponding to a given value of $i$.  The axes of these helices pass through the vertices of a honeycomb lattice.  This is a bipartite lattice, and the helices with axes on nearest-neighbor vertices have opposite handedness. For the level-1 lattice, there are two possible chirality patterns. Level 1 can form such that the light gray (red) helices in Fig.~\ref{fig:chirality_surroundings}(a) are either right-handed or left-handed. The dark (blue) helices and light (red) helices have opposite chiralities. The chirality pattern of level-1 fixes that of the higher levels: the chirality of a level-$n$ helix is opposite to that of the level-($n-1$) helix which it surrounds, for $n>1$, as is depicted in Fig.~\ref{fig:chirality_surroundings}(b).

The level-1 lattice of helices can form such that the centers of the honeycomb cells fall on any one of the sublattices $A$, $B$, $C$, or $D$ of Fig.~\ref{fig:levels1and2order}(a).  We let $S_{1,0}$ denote this choice, where the index $1$ denotes the level and the index $0$ specifies the value of $i$ corresponding to this set of helices.

\begin{figure}[tb]
        \centering
        \includegraphics[width =0.7\columnwidth]{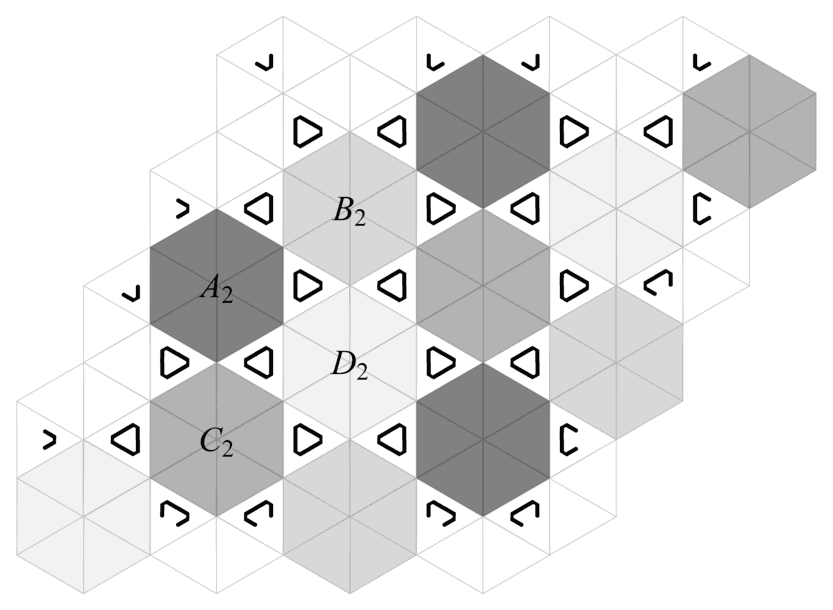}
        \caption{A 2D projection of a region of an ordered level-1 structure. The different sublattices shown in different shades of gray contain only the cubes with layer indices in $\ell^0_2$.  Bars not contributing to the level-1 structure are omitted for clarity.
        \label{fig:3D_L1}}
\end{figure}

\begin{figure*}[bt]
        \centering
        \includegraphics[width = \textwidth]{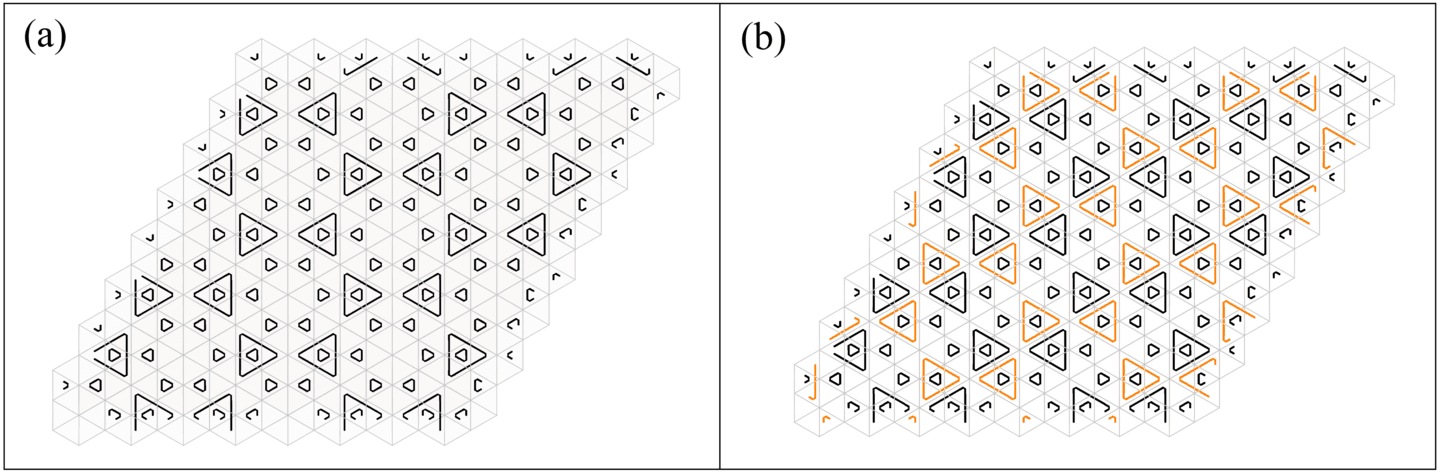}
        \caption{(Color online.)  (a) 2D projection of the two level-2 subsets ordered on the same sites. (b) 2D projection of the two level-2 subsets ordered with the centers of their honeycomb lattice on different sites, one depicted in black, the other in gray (orange).  In both (a) and (b), designs not contributing to the level-1 or level-2 structures have been left out for clarity.  
        \label{fig:3D_L2}}
\end{figure*}

Given the value of $S_{1,0}$, the level-2 honeycomb cell centers can again lie on any of four sublattices, which we denote by $S_2 \in \{A_2, B_2, C_2, D_2\}$ (see Figs.~\ref{fig:3D_L1} and~\ref{fig:3D_L2}).
Moreover, each of the two sets of helices corresponding to different values of the index $i$ defined above can have a different value of $S_2$, which we label $S_{2,0}$ and $S_{2,1}$.  Iterating this process for choosing sublattices at each scale, we see that the location of the honeycomb lattice of helices with corners in $\ell_{n}^{(i)}$ is uniquely specified by the sequence $\{S_{1,0}, S_{2,s_2}, S_{3,s_3},\ldots , S_{n,s_n}\}$, where $s_{n} = i$ and $s_{n'} = \mod_{2^{n'-1}} s_{n'+1}$ for $1 \leq n' < n$.  
Recalling that $i$ can take any of $2^{n-1}$ values, we find that specifying a fully ordered level-$n$ structure requires specifying $2^{n}-1$ values $S_{n,i}$, yielding a degeneracy 
\begin{equation}
g_n  = 2\times 4^{(2^n-1)}\,,
\end{equation}
where the factor of 2 accounts for the two possible chirality patterns of level 1.

We prove in the Appendix that this system has no periodic ground states.  We further conjecture that the limit-periodic states exhaust the degenerate class of ground states, but we cannot rule out the possibility of other nonperiodic states.  

\subsubsection{Thermodynamically favored states and definition of the order parameter}
Monte Carlo simulations indicate that a slow cooling of the zonohedral model produces one of the  limit-periodic states described in the previous section. These states emerge through a series of transitions corresponding to the sequential ordering of helices of the different levels.  To quantify the order arising as the system cools, a set of order parameters describing each level and each value of $i$ is required.

\begin{figure}
        \centering
        \includegraphics[width = 1.0\columnwidth]{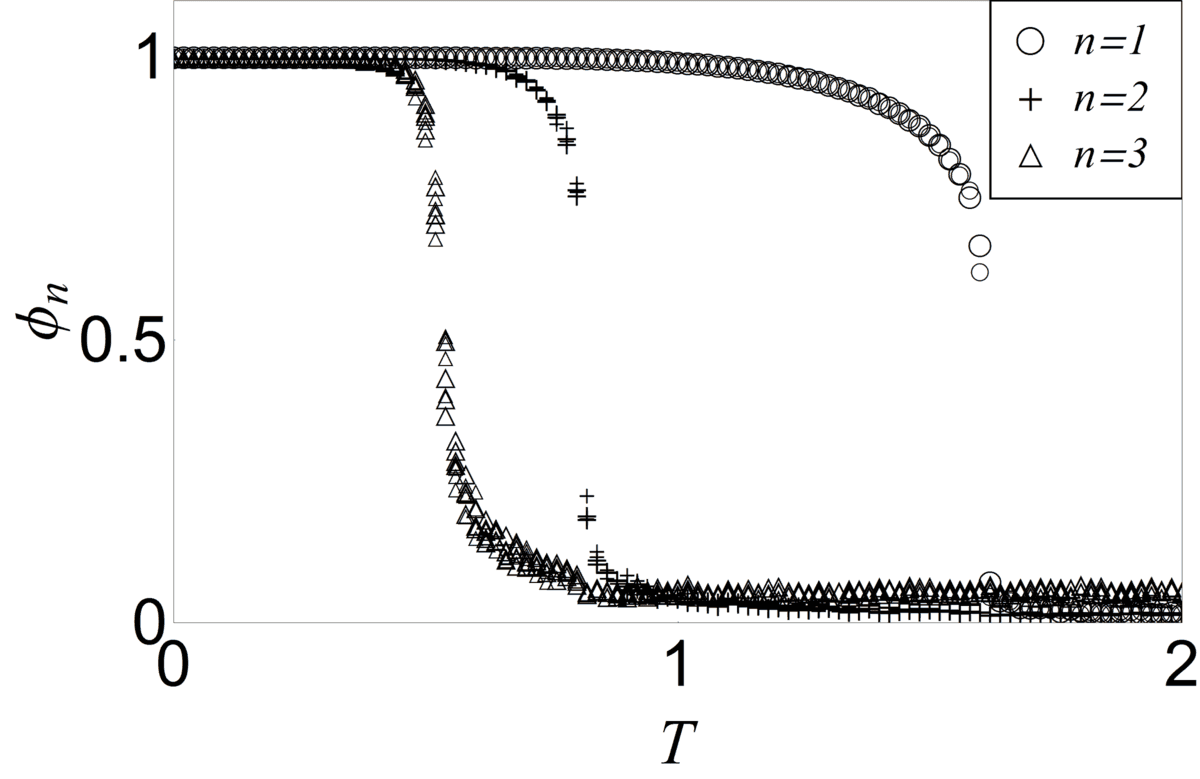}
        \caption{Order parameters of each subset of helices for the first three levels of the zonohedral model with $\e_2 = \e_1 = 1$. There is one order parameter for level-1, two for level-2, and four for level-3. The system is cooled from $T = 2$ to $T = 0$, in increments of $\Delta T = .02$ with $\tau = 1\times10^5$ MCS, then heated in the same manner. Simulations are performed on a rhombic lattice of size $16\times16\times24$. During the cooling process, the order parameter of a level-$n$ subset is found only after level-$(n-1)$ is ordered. The order parameters do not go to zero at high temperatures because of finite-size effects. 
        \label{fig:order11}}
\end{figure}

On each subset of level-$n$ sublattices defined in Section \ref{sec:groundState},  we define a staggered tetrahedral order parameter as in Section \ref{sec:orderParameters}. Throughout the simulations, a distinct order parameter was calculated for each subset of helices. The order parameters as a function of temperature for the seven subsets of helices of levels 1, 2, and 3 are shown in Fig.~\ref{fig:order11}.
The figure shows both heating and cooling sweeps.  The fact that the curves coincide quite closely suggests that the phase transition is second order, but it is difficult to rule out the possibility of a weakly first order transition.  In fact, the following section presents strong evidence for a first order transition in the case $\e_2 = 0$.  We conjecture that the transition becomes first order for any $\e_2 < 1$, but a full investigation of this point is beyond the scope of this work.

We have also measured the two-point correlations of $\phi_1$ to see whether there is any significant anisotropy in the development of the ordered phase.  At temperatures just above $T_{c;1}$, we find that the decay lengths for correlations in the plane and correlations along the $c$-axis are roughly equal when the geometry of the model is taken to correspond to the zonohedral unit cell of the FCC lattice (as in Fig.~\ref{fig:tilingmodels}(C)), which is the choice for which all nearest-neighbor interactions have the same bond length.  The formation of helices that project onto 2D triangles proceeds in tandem with the formation of the lattice of triangles in any given in-plane layer.

\subsection{The cubic model: $\e_2 = 0$}
\label{sec:cubic}
The possible structures of the cubic model with $\e_2 = 0$ include all of the limit-periodic states described in the previous section, a large class of periodic states, and possibly others.  In the following, we set $\e = \e_1$, yet here again we find through Monte Carlo simulations that the thermodynamically favored states are the limit-periodic states described in the previous section. When the system is slowly quenched, the ground state is reached in a similar manner; i.e.,  through a series of phase transitions corresponding to the ordering of the level-$n$ helices.

\begin{figure}
        \centering
        \includegraphics[width = 1.0\columnwidth]{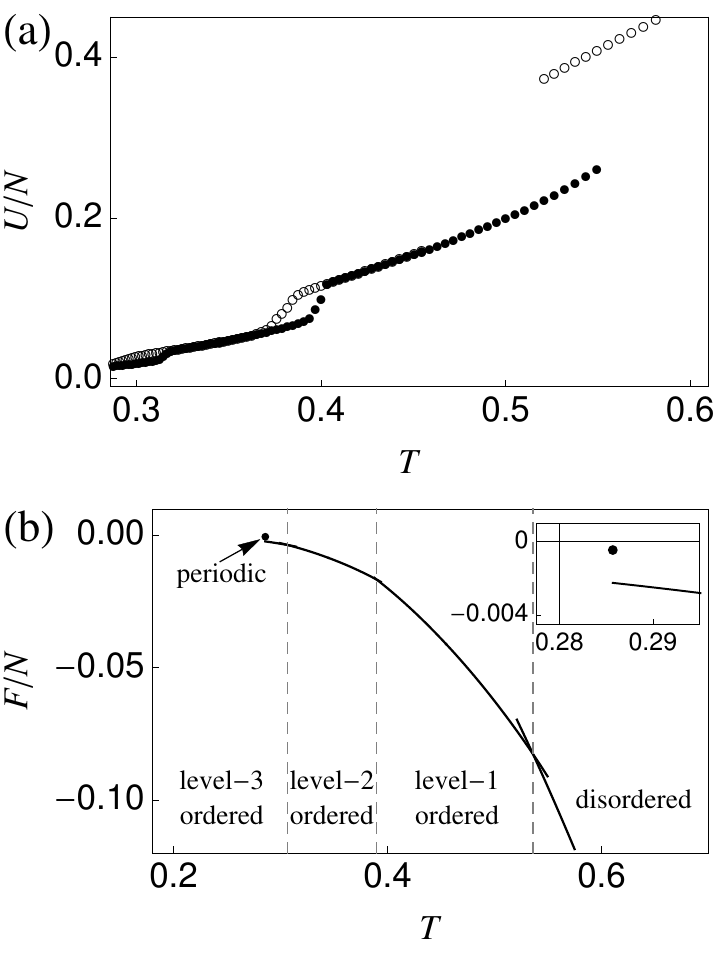}
        \caption{(a) Energy per site during a slow quench and subsequent heating. The phase transitions correspond to those of levels 1, 2, and 3, in order of decreasing $T$.  See the end of Section~\ref{sec:cubic} for the details of the simulation. (b) Free energy of the disordered, level-1 ordered, level-2 ordered, and level-3 ordered phases.  The dot indicates the free energy of the simplest periodic competing phase.
        \label{fig:UandF}}
\end{figure}

The phase transitions, however, are now clearly first order. The energy curves as a function of temperature exhibit hysteresis, as can be seen in Fig.~\ref{fig:UandF}(a).  Though the scaling argument used for the 2D case still holds, the scaling collapse is difficult to observe because the size of the hysteresis loop observed in numerical simulations depends on the rate of cooling or heating, and we do not know how to scale those rates to achieve a clean collapse.   We therefore carefully study the nature of the transition and the scaling by computing the relevant free energies.  The free energies of the different phases as a function of temperature (computed using a technique described below) show clear discontinuities in slope within the hysteresis loop, further verifying the order of the transition.

\subsubsection{Free energy calculations}
\label{sec:freeenergy}
Free energies of the cubic model can be computed as follows.
Let $N$ be the number of lattice sites in the system, $u$ be the internal energy per site, and $f$ be the Helmholtz free energy per site. The fundamental thermodynamic identity and the definition of Helmholtz free energy imply the following relationship between $u$ and $f$:
\begin{equation}
   f(\beta_1) = \frac{1}{\beta_1} \left[ \beta_0 f(\beta_0) + \int_{\beta_0}^{\beta_1}u \, d\beta \right]\,,
\label{eq:Fb} 
\end{equation}
where $\beta \equiv 1/T$ and $\beta_0$ and $\beta_1$ are fixed inverse temperatures.

Evaluating the right-hand side of Eq.~\eqref{eq:Fb} requires independent knowledge of the value of $\beta_0 f(\beta_0)$ for a temperature range at which the phase under consideration is stable. We study four phases: the disordered state, the state in which level 1 is ordered, one in which levels 1 and 2 are ordered, and one in which levels 1, 2, and 3 are ordered.  We refer to a state in which all levels up to and including level $n$ are ordered as the ``level-$n$ ordered state.''

The calculation of $\beta_0 f(\beta_0)$ for the disordered state is easily done for $\beta_0 = 0$.  Let the internal energy at $\beta_0 = 0$ be $u_0$. Because there are six possible orientations per site, the entropy is:
\begin{equation}
        s_0 \equiv \frac{S_0}{N} =  \ln 6\,.
\end{equation}
Because $\beta_0 u_0 = 0$ and the free energy $f = u - s/\beta$, we have
\begin{equation}
        \lim_{\beta_0 \rightarrow 0} \beta_0 f(\beta_0) = -\ln 6\,.
\end{equation}

To determine $\beta_0 f(\beta_0)$ for a level-$n$ ordered state at an appropriate value of $\beta_0$, we use thermodynamic integration~\cite{FrenkelBook}. This method consists of finding some reference system for which the free enery can be determined analytically and from which there is a smooth path in parameter space to the system of interest (not passing through any phase transitions).  The system is then monitored during a simulation in which a parameter is slowly varied, which switches the Hamiltonian from that of the reference system $H_0$ to that of the system of interest $H_1$.  Let the Hamiltonian $H_{\lambda}$ be
\begin{equation}
        H_{\lambda} \equiv (1-\lambda)H_0 - \lambda H_1\,,
        \label{Hl}
\end{equation}
where $\lambda$ ranges from 0 to 1.  The free energy $f(\beta_0)$ of the system of interest is computed using the relation:
\begin{equation}
 f(\beta_0) - f_0(\beta_0) = \frac{1}{N}\int_0^1 \bigg \langle \frac{\partial H_{\lambda}(\beta_0)}{\partial \lambda} \bigg \rangle\, d\lambda\,,
 \label{eq:frenkel}
\end{equation}
where $f_0(\beta_0)$ is the free energy for the $\lambda=0$ Hamiltonian $H_0$.

In the cubic model, $H_1$ is the sum of the interaction energies of all the sites in the lattice with their nearest neighbors.  The reference Hamiltonian, $H_0$, is a sum of two conjugate fields interacting with subsets of the lattice sites.  One of the fields interacts with the cubes that form the corners of the desired levels, while the other interacts with the edges.  The cubes do not interact with each other.

The free energy of the reference system of ordered levels with index less than $n$ is calculated as follows.  Define on each lattice site an integer-valued pseudospin, $q$, with $1 \le q \le 6$, corresponding to one of the six configurations of the tile, and split the system into three non-interacting systems:
\begin{description}
        \item[system 1] a system of non-interacting, free spins
        \item[system 2] a paramagnetic system consisting of all tiles contributing edges to the ordered levels with index less than $n$
        \item[system 3] a paramagnetic system consisting of all tiles contributing corners to the ordered levels with index less than $n$, but not contributing edges to any of those levels.
\end{description}
The calculation of the free energy of system 1 is straightforward. A structure with ordered levels of index less than or equal to $n$ leaves $N/4^n$ cubes unrestricted. Each of these cubes has six equally probable spins.  Because the spins do not interact with any part of the system, their internal energy is 0. The free energy per site is then:
\begin{equation}
        f_{0} = -\frac{S}{N \beta} = -\frac{1}{4^n \beta}\ln 6\,.
\end{equation}

The free energy of system 2 is calculated in the presence of a conjugate field $h_i(q_i)$ that takes the value 0 for all $q_i$ consistent with an ordered state, and $h_0>0$ for all $q_i$ inconsistent with the ordered state. An edge is defined by the location and orientation of the black bar on a cube. The specification of an edge uniquely defines the configuration of a tile. Thus, there is one value of $q$ for which $h(q)$ is 0, and five for which $h(q)$ is $h_0$. The total Hamiltonian of the edge system is:
\begin{equation}
        H_{\mathrm{E}} =  \sum_{i=1}^{N_\mathrm{E}} h_i(q_i),
        \label{eq:H1}
\end{equation}
where $N_E$ is the total number of edges of levels with index less than $n$. The total number of edges in level-$n$ is the length of an edge ($2^{n-1}-1$)  times the number of edges (3) times the total number of level-$n$ triangles in the system ($2N/4^n$):
\begin{equation}
        N_{\mathrm{E}} = N\sum_{i=1}^n 3(2^{i-1}-1)\frac{2}{4^i}\equiv N n_{\mathrm{E}}\,,
\end{equation}
where $n_{\mathrm{E}}$ is the fraction of spins in system 2. The partition function is 
\begin{eqnarray}
        Z_{\mathrm{E}} & =  &\sum_{\mathrm{all~configs.}} e^{-\beta \sum_i  h_i(q_i) }  \\
        \ & = & \prod_{i=1}^{N_{\mathrm{E}}} \sum_{q_i} e^{-\beta h_i(q_i)} \\
        \ & = & \l( 1 +  5e^{-\beta h_0 }\r)^{N_{\mathrm{E}}}\,,
        \label{eq:Z1}
\end{eqnarray}
which yields the free energy:
\begin{equation}
        f_{\mathrm{E}} = -\frac{n_E}{\beta}\ln\l(1+5e^{-\beta h_0}\r)\,.
\end{equation}

System 3 can be treated in a similar manner, except that the degeneracies of the individual cube energy states are now different. The definition of the orientation and location of the corners of a cube does not uniquely specify the configuration, but does restrict it to two possibilities. Therefore $h_i(q_i)$ gives 0 for two spin values and $h_0$ for four spin values. The total number of corner cubes in the structure of level-$n$ is:
\begin{equation}
        N_{\mathrm{C}} = N - N_{\mathrm{E}} - \frac{N}{4^n} \equiv N n_{\mathrm{C}}\,.
\end{equation}
The partition function is
\begin{equation}
        Z_{\mathrm{C}}  =  \l( 2 +  4e^{-\beta h_0 }\r)^{N_{\mathrm{C}}}\,,
        \label{eq:Z3}
\end{equation}
and the free energy is
\begin{equation}
        f_{\mathrm{C}}  = -\frac{n_{\mathrm{C}}}{\beta}\ln\l(2+4e^{-\beta h_0}\r)\,.
\end{equation}
The total free energy of the complete reference system of ordered levels with index less than or equal to $n$ is:
\begin{eqnarray}
        f_{n,\mathrm{ref}} & = & -\frac{1}{\beta}\Big[ 4^{-n}\ln 6 +  n_{\mathrm{C}}\ln \l(2+4e^{-\beta h_0} \r)  \nonumber \\
        \ & \ & \quad\quad\quad  + \,  n_E \ln\l(1+5e^{-\beta h_0}\r)\Big].
\end{eqnarray}

The simulation parameters used for the thermodynamic integrations are listed in Table~\ref{tab:param}. Integration of $\langle \partial H_{\lambda}/\partial{\lambda} \rangle$ as a function of $\lambda$ is performed using Gauss-Lobatto quadrature with 20 abscissas. The results are plotted in Fig.~\ref{fig:UandF}(b). The discontinuities in slope of the free energy curves indicate that the phase transitions are first order.  The critical temperatures of the first three transitions can be determined by locating the crossings of the curves obtained for the different phases and are presented in Table~\ref{tab:Tc}. 
\renewcommand{\arraystretch}{1.5}
\begin{table}[tb]
        \centering
        \begin{tabular}{| c || c | c | c | c|}
                \hline
                  &        level 1 & level 2 & level 3 & periodic\\
                \hline
                \hline
                 Size &  $8\!\times\!8\!\times\!12$ & $16\!\times\!16\!\times\!24$ & $32\!\times\!32\!\times\!48$ & $32\!\times\!32\!\times\!48$\\
                \hline
                 $ \beta_i$& 0 & 2.2 & 2.9 & 3.5\\
                \hline
                 $ \beta_f$& 2.2 & 2.9 & 3.5 & N/A\\
                \hline
                 $\Delta \beta$ & .02 & .02 & .02 & N/A\\
                \hline
                 $\tau_{\beta}$ & $10^5 $& $10^5 $& $10^5 $ & $10^5$\\
                \hline
        \end{tabular}
        \caption{Monte Carlo parameters used for obtaining the internal energy for levels 1, 2,  and 3 of the limit-periodic sequence and for a system prepared in the 3-periodic state of Fig.~\ref{fig:periodic_image}. $\tau_{\beta}$ is the number of Monte Carlo steps performed at each temperature.}
        \label{tab:param}
\end{table}

\renewcommand{\arraystretch}{1.5}
 \begin{table}[tb]
     \centering
     \begin{tabular}{|c|}
            \hline 
             $T^*_{c;1}$ = 0.5359 $\pm\,1.3\times10^{-3}$ \\
             \hline
             $T^*_{c;2}$ = 0.3898 $\pm\,6\times10^{-4}$\\
             \hline
             $T^*_{c;3}$ = 0.3065 $\pm\,3\times10^{-4}$\\
            \hline
     \end{tabular}
        \caption{Critical temperatures of the transitions for the first three levels in the cubic model.}
        \label{tab:Tc}
   \end{table}

\begin{figure}[tb]
        \centering
        \includegraphics[width = 1.0\columnwidth]{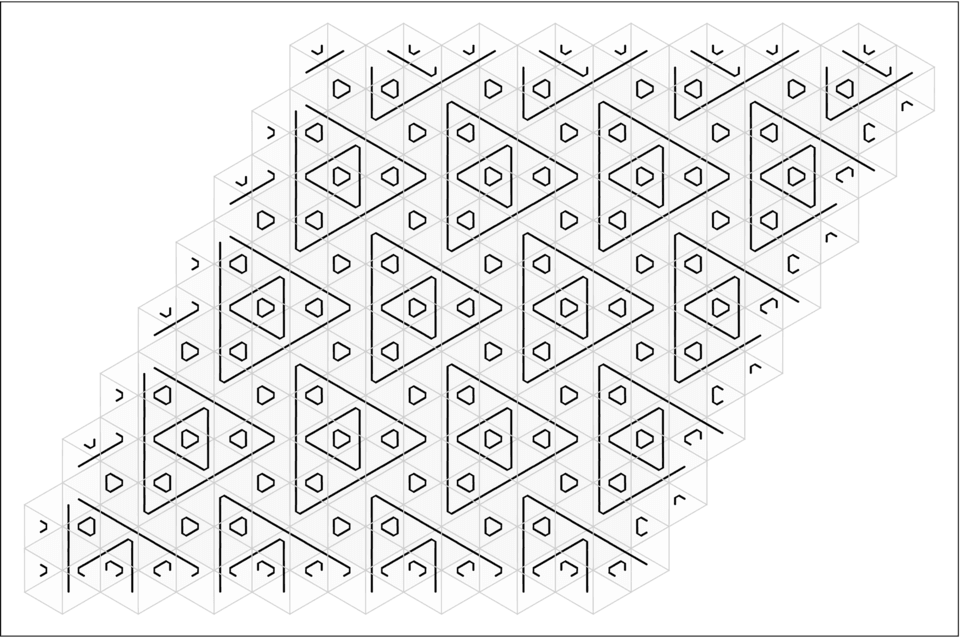}
        \caption{Section of one of the periodic ground states. The largest helices belong to level 3.
        \label{fig:periodic_image}}
\end{figure}

We have also computed the free energy of the simplest low energy periodic phase at a temperature where it might be expected to compete with a state in the limit-periodic hierarchy -- just below $T^*_{c;3}$. The  zero-energy periodic state with the smallest unit cell is shown in Fig.~\ref{fig:periodic_image}.  Based on the size of the largest triangles in this state, we refer to it as the ``3-periodic'' structure. 
The reference Hamiltonian used here is a single conjugate field interacting with all the tiles in the system. As in system 2, at each site, $i$, one value of the pseudospin, $q_i$, yields $h_i(q_i) = 0$ while five values yield $h_i(q_i) = h_0$. Thus, the free energy per tile in the reference system is:
\begin{equation}
        f = -\frac{1}{\beta}\ln(1+5e^{-\beta h_0})\,.
\end{equation}

We find that the 3-periodic state is metastable; its free energy per tile at $\beta=3.5$ is $-4.69\times10^{-4}$, which is clearly higher than that of the competing level-3 ordered state in the limit-periodic hierarchy.   The internal energy of the 3-periodic state is lower than that of the level-3 ordered state by approximately $0.011$ per tile, but the level-3 state has the higher entropy due to the fluctuations of the level-4 corners and edges.  In particular, the lack of edges longer than 3 tiles significantly suppresses the entropy of the 3-periodic state.  The difference in the free energies of these two phases is approximately $0.002$ per tile, corresponding to an energy cost of one mismatch per 500 tiles, or roughly one mismatch per 10 unit cells of the 3-periodic structure.  

\subsubsection{Scaling relations}
\label{sec:freeenergyscaling}
A scaling argument similar to the one discussed in Section~\ref{sec:scalingRelations} applies to the limit-periodic structures formed by the cubic model as well.  A configuration of this system is specified by giving the location of the ends of the black bars on each of the faces. A configuration is allowed if the specification of the positions of these objects for every tile corresponds to a possible orientation of either of the two enantiomorphs shown in Fig.~\ref{fig:tilingmodels}(d).
Suppose that level~$(n-1)$ is completely ordered and all decorations of cubes forming the helices of all levels with index less than $n$ are fixed. The level-$n$ system is defined to include all of the remaining degrees of freedom: (1) the set of tiles left unused when level~$(n-1)$ is ordered and (2) the long bars that form bonds of length $2^{n-1}-1$ between these tiles. The long bars in (2) are on tiles that form the corners of the levels with index less than $n$. These bonds could in principle form the edges of the helices.

Within the level-$n$ system, there are $2^{n-1}$ non-interacting subsystems. An individual subsystem will be referred to as level-$n_i$, where $i$ specifies the layers, $\ell_n^{(i)}$, on which the centers of the unused tiles are located. This subset of unused tiles is defined as $U_{n,i}$.

\begin{figure}[tb]
        \centering
        \includegraphics[width = 0.5\columnwidth]{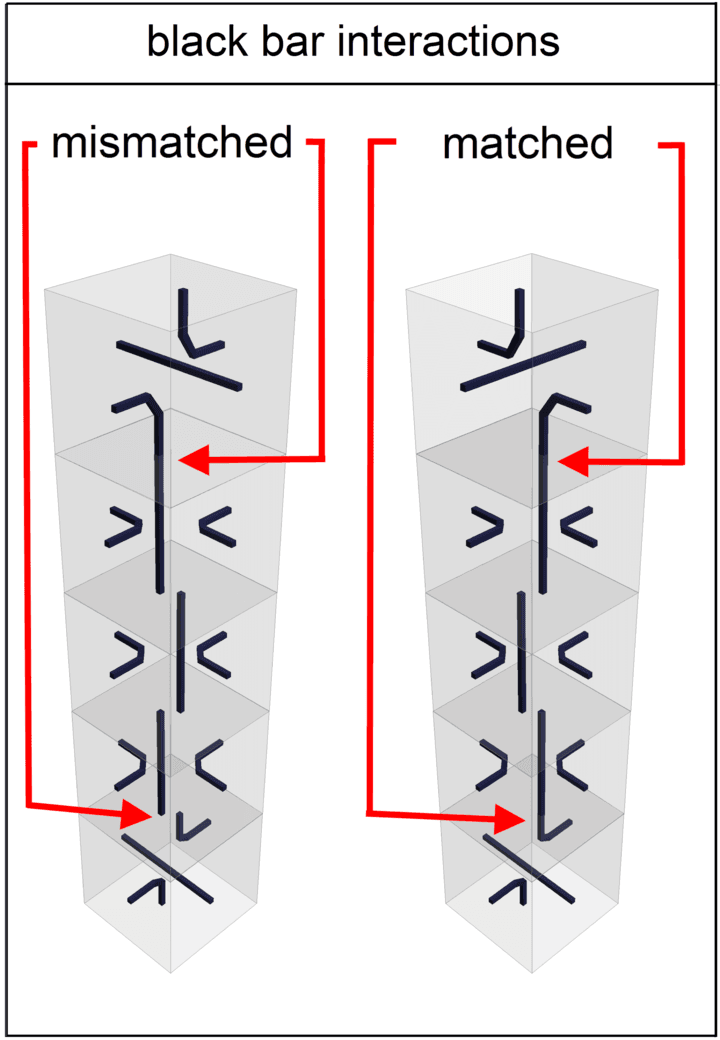}
        \caption{Matched and mismatched corner configurations for the black bonds of level-3.  
        \label{fig:matched}}
\end{figure}

For a given configuration of tiles in $U_{n,i}$,  the partition function of the level-$n_i$ subsystem is the product of the individual level-$n$ bond partition functions, $\zeta_n^{\pm}(T_n)$. Thus, the partition function of the level-$n_i$ subsystem is:
\begin{equation}
        Z_n(T_n) = \sum_{\mathrm{config.}}\prod_{\mathrm{n.n}}\zeta^{\pm}_{n}(T_n)\,, 
        \label{eq:z}
\end{equation}
where the partition function of an individual \mbox{level-$n$} bond with matching (mismatching) corners is $\zeta^{+}_n$ ($\zeta^{-}_n$).  $\zeta^{\pm}_n$ is found by treating the level-$n$ bonds as 1D Ising chains with $k_n \equiv 2^{n-1}$ spins, and the same scaling relations of Eq.~\eqref{eq:nbpart} are obtained.  Drawings of matched and mismatched corners for this case are shown in Fig.~\ref{fig:matched}. 

The configuration sum of the subsystem of tiles in $U_{n,i}$ is identical to that of the system of tiles in $U_{1,0}$, which is the set of all tiles.  Both describe cubic lattices with bonds joining neighbors along the principal axes directions.  In exact analogy with the 2D models, there is a temperature $T_n$ at which the level-$n_0$ system behaves identically to the level-$1_0$ system at $T_1$, which implies
\begin{equation}
        Z_n(T_n) = A_n Z_1(T_1)\,
\end{equation}
for some constant $A_n$.
The relation of $T_n$ to $T_1$ is determined by equating the partition functions for individual bonds:
\begin{align}
        \zeta^{+}_n(T_n) &= \alpha_n \zeta^{+}_1(T_1)\\
        \zeta^{-}_n(T_n) &= \alpha_n \zeta^{-}_1(T_1)\notag\,,
\end{align}
where $\alpha_n = A_n^{1/N_b}$, with $N_b$ being the number of bonds.  As for the 2D case, we get:
\begin{equation}
        \tanh\l(\frac{\e}{2 T_1}\r) = \l[\tanh\l(\frac{\e}{2T_{n}}\r)\r]^{k_n}\,.
        \label{eq:scaling3D}
\end{equation}
or, equivalently, 
\begin{equation}
        \tanh\l(\frac{\e}{2 T_n}\r) = \l[\tanh\l(\frac{\e}{2 T_{n+1}}\r)\r]^2\,.
        \label{eq:scalingn}
\end{equation}

The complete level-$n$ system includes $k_n$ independent level-$n_i$ subsystems. Therefore, the partition function of the full level-$n$ system is:
\begin{equation}
        Q_n(T_n) =  \l[Z_n(T_n)\r]^{k_n}\,,
\end{equation}
where $Z_n$ is defined in Eq. \eqref{eq:z}.  The free energy of the level-$n$ system is:
\begin{eqnarray}
F_n(T_n) & = & - T_n \ln Q_n (T_n) \nonumber \\
      & =  & k_n\left[ F_1(T_1) - T_n \ln A_n\right]\,,
        \label{eq:fnScaled}
\end{eqnarray}
where $A_n = (\zeta_n^{\pm}(T_n)/\zeta_1^{\pm}(T_1))^{N_b}$.

Because the transition temperatures are finite, the assumption that all levels with indices less than $n$ are completely fixed is not strictly satisfied, as in the 2D case. At the \mbox{level-$n$} transition, tiles in levels with index less than $n$ may fluctuate.  A straightforward correction to the free energy derived above can be made by considering the edges of levels with indices less than $n$ for $n>2$. Fluctuations of these edges have no effect on the bonds between \mbox{level-$n$} corners, and thus have no effect on the scaling argument for $T_n$.  To calculate this correction to the free energy of the \mbox{level-$n$} system, consider the edge of a \mbox{level-$m$} triangle, where $m<n$. In each edge consisting of $2^{m-1}-1$ long bars, there are $k_m=2^{m-1}$ bonds. Each bond can either be in a matched state with energy 0 or a mismatched state with energy $\e$.  In a single edge, there must be both an even number of mismatches and an even number of matches. The number of \mbox{level-$m$} edges in a system with $N$ tiles, $N_{m,E}$ is:
\begin{equation}
        N_{m,E} = \frac{6N}{4^m}\,.
\end{equation} 
Thus, the partition function of the system of \mbox{level-$m$} edges at a temperature $T_n$  is:
\begin{equation}
        z_m(T_n) = \left[\sum_{i=0}^{k_m/2}\binom{k_m}{2i} e^{-2i \e /T_n}\right]^{6N/4^m}.
\end{equation}

The partition function of the level-$n$ system including the fluctuations on edges of levels with index less than $n$ is:
\begin{equation}
     Q_n^{\prime}(T_n)  =  Q_n(T_n)\prod_{m=2}^{n-1} z_m(T_n) \,.
\label{eq:Qnprime}
\end{equation}
Equations~\eqref{eq:fnScaled} and~\eqref{eq:Qnprime} yield the free energy:
\begin{equation}
      F_n^{\prime}(T_n)  =   k_n F_1(T_1) - T_n\left[k_n \ln A_n + \sum_{m=2}^{n-1} \ln z_m(T_n)\right],
\label{eq:f3Scaled}
\end{equation}
where $T_n$ is related to $T_1$ through Eq.~\eqref{eq:scaling3D}.
Note, however, that this expression does {\em not} account for fluctuations in the corners of the (ordered) lower levels.  Thus, as in the 2D case, the prediction of Eq.~\eqref{eq:f3Scaled} is not exact.

\begin{figure}[tb]
        \centering
        \includegraphics[width = 1.0\columnwidth]{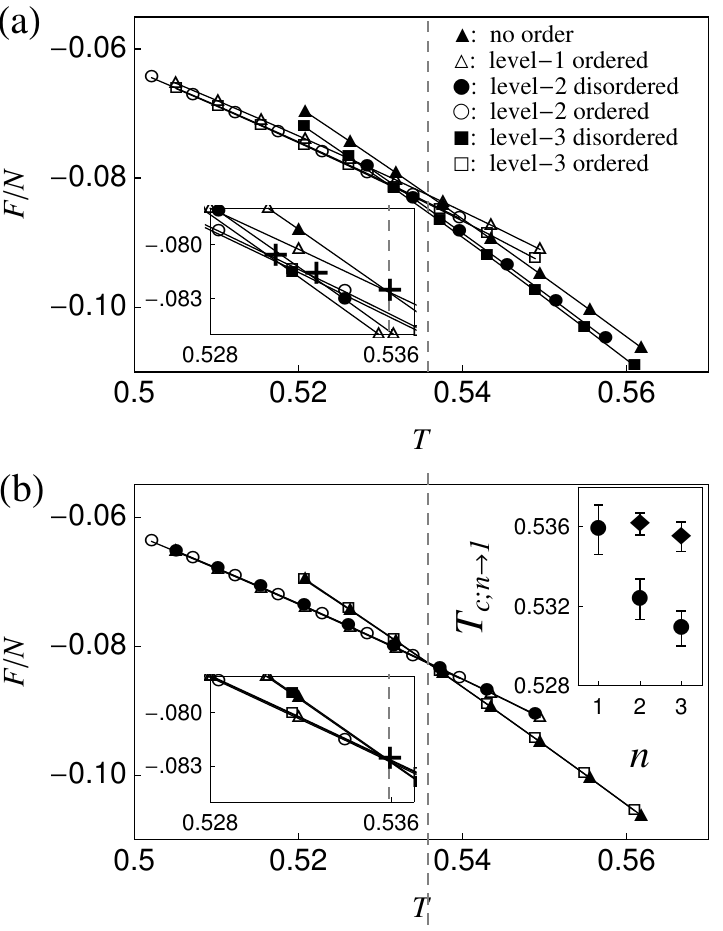}
        \caption{
        (a) Free energy of levels 2 and 3 scaled onto that of level-1 according to Eq.~\eqref{eq:fnScaled} in the full system (fluctuations are {\em not} artificially suppressed. Level-2 and level-3 curves have been  scaled according to Eq.~\eqref{eq:fnScaled}.  The scaled transition temperatures for the different levels, marked by the heavy $+$’s, are shown in the inset of panel (b).
        (b) Free energy of levels-2 and 3 scaled onto that of level-1 according to Eq.~\eqref{eq:fnScaled} when levels with indices less than $n$ are held fixed, showing perfect agreement with the scaling theory.  The vertical dashed line through the intersection of the curves obtained by integrating down from high $T$ and integrating up from low $T$ marks the transition temperature.  
        Inset: Critical temperatures of the levels-2 and 3 transitions scaled onto that of level-1. The diamonds are determined from simulations in which levels with indices less than $n$ are fixed by a conjugate field as in (b). The disks correspond to the systems with all equilibrium fluctuations present.
         \label{fig:freeEnergy}}
\end{figure}

The validity of the scaling relations was tested by applying them to the free energy curves and critical temperatures obtained from simulations. Using Eq.~\eqref{eq:scaling3D}, the predicted transition temperatures of \mbox{level-2} and \mbox{level-3} from the value of $T^*_{c;1}$ in Table~\ref{tab:Tc} are 0.3910 $\pm\,7\times10^{-4}$ and 0.3082 $\pm\, 4\times10^{-4}$, respectively, corresponding to $0.5\%$ relative error. The scaling theory, however, assumes that all bonds in levels with indices less than $n$ are fixed.  This would imply that $\phi_{n-1}$ is strictly equal to unity in the vicinity of $T_{c;n}$, but the actual value of $\phi_1$ at $T_{c;2}$ is measured to be 0.999.  To check the scaling theory and our numerical determinations of the free energy, we perform simulations in which levels with indices less than $n$ were fixed by hand. The excellent agreement between scaling predictions and simulations is shown in Fig. \ref{fig:freeEnergy}(a).  

The perfect scaling collapse of the free energy curves of the \mbox{level-3} and \mbox{level-2} systems onto that of \mbox{level-1} when the relevant lower levels are fixed externally indicates that the source of any deviations from the scaling relation is the fluctuations of the corner tiles in lower levels.  Figure~\ref{fig:freeEnergy}(a) gives an indication of the size of those deviations, and the inset of Fig.~\ref{fig:freeEnergy}(b) shows the results for scaled transition temperatures computed from simulations.  These results strongly suggest that the accuracy of the scaling argument improves as $n$ increases, implying that the infinite sequence of transitions is not disrupted by the cumulative effect of residual fluctuations in each layer.

%
\section{Temporal scaling and kinetic barriers}\label{sec:barriers}
As noted in Ref.~\cite{Byington2012}, the hierarchy of phase transitions can lead the system to fall out of equilibrium when quenched too rapidly.  Roughly speaking, simultaneous attempts to establish order at two or more levels creates a competition resulting in defects that require extremely long times to heal due to their complex geometric and topological structures.  This raises two questions. (1) How slow does a quench have to be in order for the limit-periodic state to be accessed?  And (2) what is the nature of the defects that prevent equilibration when the quench is too rapid?  We consider these questions here in the context of the 2D Taylor-Socolar lattice model with $\e_2 = \e_1$.

\subsection{Relaxation times for ordered phases}
As the temperature is lowered, full ordering requires that the level-$n$ order be firmly established before the critical temperature for level-$(n+1)$ is reached.  Consider a cooling protocol in which the temperature is varied in a sequence of steps, being fixed at temperatures $T_{q;n}$ for a time $t_n$, where $T_{q;n}$ lies between $T_{c;n}$ and $T_{c;n+1}$.  We take the $T_{q;n}$’s to be related by the scaling relation of Eq.~\eqref{eq:scaling}.  Our goal is to find the minimal values of $t_n$ such that $\phi_n$ reaches its equilibrium value before the temperature is lowered, which requires understanding how the times for relaxation to equilibrium scale with $n$ for large $n$.

Because the ordering at higher levels requires the equilibration of longer bonds between triangle corners, we expect $t_n$ to increase with $n$.  A first estimate of the scaling of $t_n$ with $n$ for large $n$ can be made based on the fact that, given complete ordering of all levels less than $n$, the level-$n$ system is identical to the level-$(n-1)$ system except that the length of the bonds between triangle corners is twice as large.  Each such bond behaves as an Ising chain that mediates the interaction between corners.  Thus the time required for equilibration should scale like the time required to establish correlations on the order of the bond length.  The correlation length for an Ising chain grows like the square root of time~\cite{Bray1990}, so we expect the time, $\tau_n$, required for corners to become correlated to scale like the square of the distance between them: $\tau_n \sim 2^{2n}$ at large $n$.  Corrections to this asymptotic form for small $n$ can be found by numerical simulation of short Ising chains.  We find, however, that the scaling argument gives a poor account of the relaxation times observed in Monte Carlo simulations of the full system.   Here we content ourselves with reporting the results of those simulations.

We choose $T_1=1.0$ to be the temperature for equilibrating the level-1 ordered phase of the Taylor-Socolar tiling model after a sudden quench from infinite temperature.  We study the equilibration of level $n$ by fixing levels $1$ through $n-1$ in their perfectly ordered states and quenching the remaining degrees of freedom from infinite temperature to the temperature $T_n$ related to $T_1$ by Eq.~\eqref{eq:scaling}.

\begin{figure}[tb]
\begin{center}
\includegraphics[width = 1.0\columnwidth]{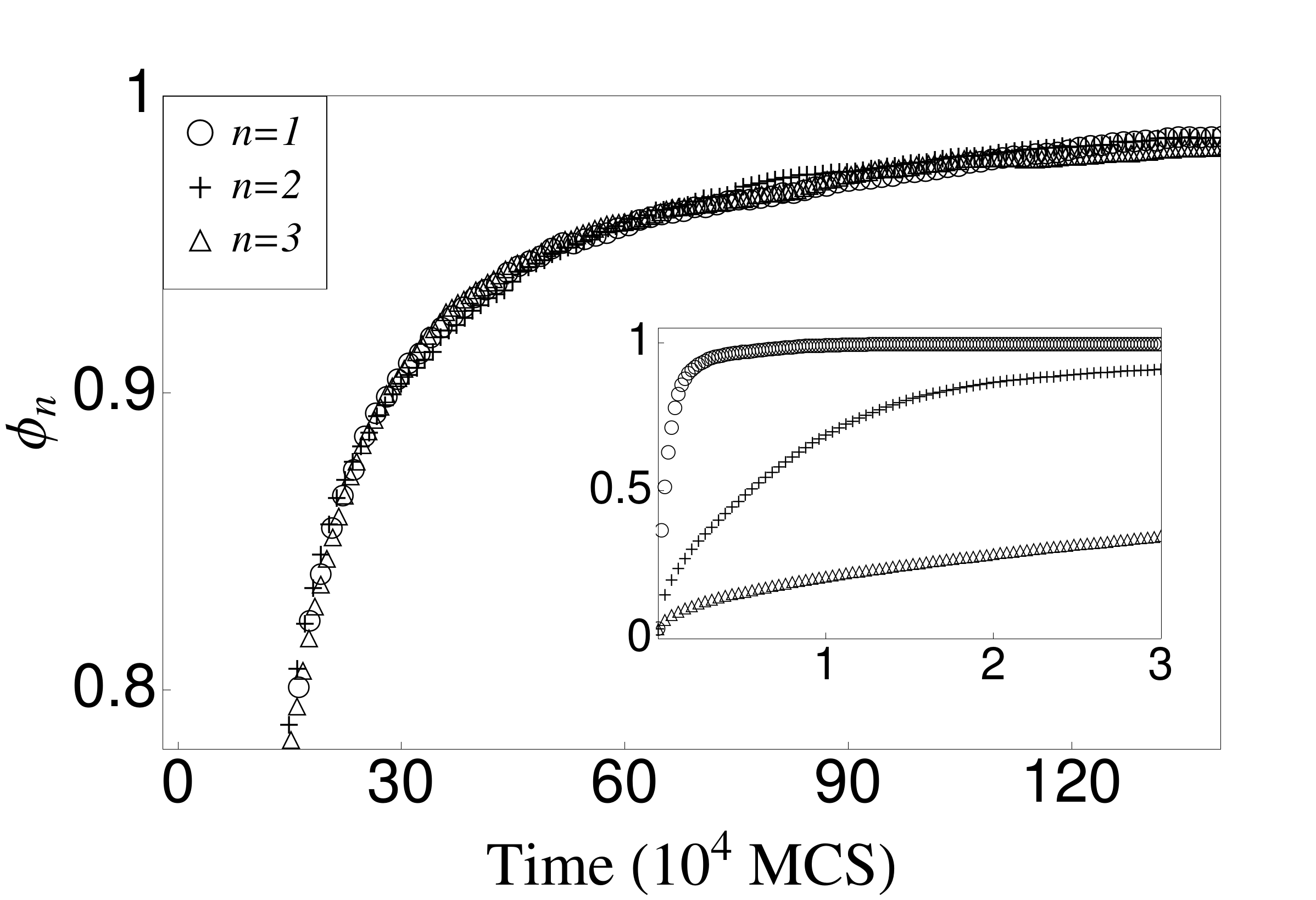}
\caption{Inset:  $\phi_n$ vs. time for levels $1$, $2$, and $3$. The level $1$ structure is relaxed on a $32\times32$ lattice at $T_1=1.0$; level $2$ is relaxed on a $64\times64$ lattice at $T_2=0.603$; level $3$ is relaxed on a $128\times128$ lattice at $T_3=0.427$.  Full panel: Data at long times from the runs shown in the inset with times scaled using $\tau_2/\tau_1=13.8$ and $\tau_3/\tau_2=10.7$.
\label{fig:order_relaxation}}
\end{center}
\end{figure}

The inset in Fig.~\ref{fig:order_relaxation} shows $\phi_n$ as a function of the number of Monte Carlo steps for $n=1$, $2$ and $3$. The lattice sizes used are $32\times32$, $64\times64$, and $128\times128$ respectively, so that the number of corner tiles to be ordered for each level is the same in all cases. Level $1$ is equilibrated at $T_1=1.0$; levels $2$ and $3$ are equilibrated at the corresponding temperatures $T_2=0.603$ and $T_3=0.427$ respectively.   From the long-time behaviors, where $\phi_n > 0.8$, we fit the relaxation to $\phi\approx 1$ with an exponential and extract the time constant $\tau_n$ listed in Table~\ref{tab:tau_ratios}.   The full panel in Fig.~\ref{fig:order_relaxation} shows the curves from Fig.~\ref{fig:order_relaxation} with the times scaled by $\frac{\tau_n}{\tau_1}$.

\begin{table}
\centering
\begin{tabular}{cc}
\hline \hline
& ratio of relaxation constants \\
\hline
$\tau_2/\tau_1$ & 13.8 \\
$\tau_3/\tau_2$ & 10.7 \\
\hline \hline
\end{tabular}
\caption{The ratios between the relaxation constants for levels $1$, $2$, and $3$ obtained from Monte Carlo simulation of the 2D tiling model.}
\label{tab:tau_ratios}
\end{table}

\subsection{Failure to order in rapid quenches}
\begin{figure}[tb]
\centering
\includegraphics[width = 1.0\columnwidth]{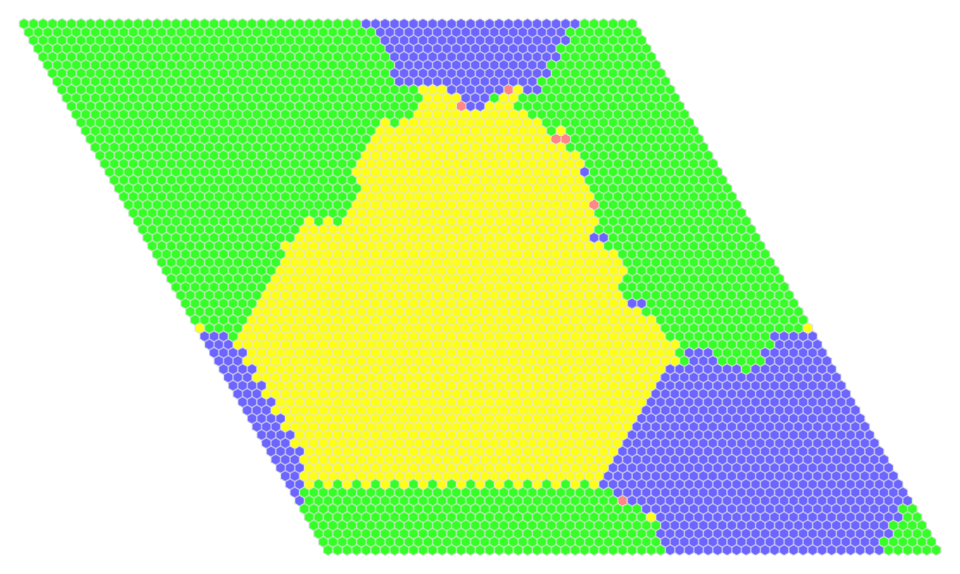}
\caption{(Color online.)  Distinct ordered regions in a configuration of a $64\times64$ lattice equilibrated for $5.4\times10^5$ Monte Carlo steps per tile after a rapid quench to $T=0.6$.  Tiles are colored according to the sublattice ($A$, $B$, $C$, or $D$) that specifies the level-1 structure.  (See Fig.~\ref{fig:levels1and2order}(a).)  The gray in the upper left (green) and dark gray (blue) regions are each single domains connected through the periodic boundary conditions.
\label{fig:domainwalls}} 
\end{figure}

In order to identify the defects that prevent equilibration in a rapid quench, we perform Monte Carlo simulations in which a random initial configuration evolves at a temperature $T=0.6$, which is below $T_{c;2}$ and above $T_{c;3}$. Figure~\ref{fig:domainwalls} shows a configuration of a $64\times64$ lattice after $5.4\times10^5$ Monte Carlo steps.  Different colors indicate different choices of sublattice for the level-1 order.    In any given region, the level-2 structure is also well ordered.

\begin{figure}[tb]
\centering
\includegraphics[width = 1.0\columnwidth]{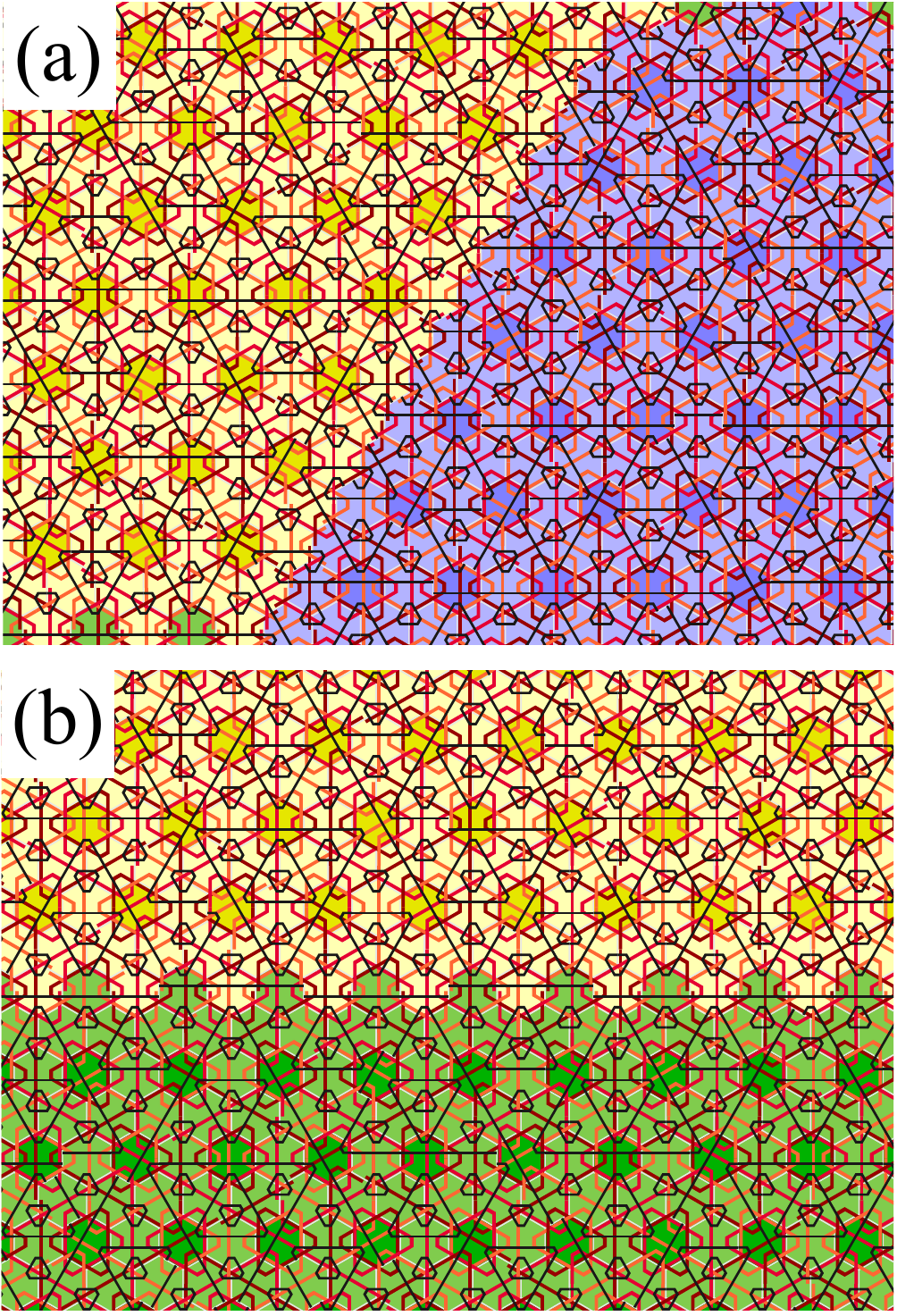}
\caption{(Color online.)  Close views of two of the domain walls formed during a sudden quench of a $64\times64$ lattice to the temperature $T=0.6$.  The top panel shows a domain wall that roughens relatively easily due to the existence of multiple mismatches along the boundary.  The lower panel shows a domain wall that becomes frozen due to the lack of black stripe mismatches in the level-1 and level-2 triangles on both sides of the boundary.
\label{fig:zoomdw}}
\end{figure}

Figure~\ref{fig:zoomdw} shows details of two types of domain walls that appear during the quench.  These are magnified images of the lower right and bottom boundaries of the central light gray (yellow) region in Fig.~\ref{fig:domainwalls}.  
The boundary in Fig.~\ref{fig:zoomdw}(a) contains multiple mismatches in the level-1 and level-2  black and thick gray (purple) structures (shown in three different colors for visual clarity), and the domain wall tends to roughen and move relatively easily.  The boundary in Fig.~\ref{fig:zoomdw}(b), however, does not contain any black stripe mismatches in either the level-1 or level-2 structure.  In this case, motion or roughening of the domain wall requires introducing multiple new mismatches in the level-1 black triangles, but because the level-2 structure is also ordered, rotations of the tiles containing level-1 corners are strongly suppressed.  These domain walls effectively block the equilibration of the level-1 and level-2 order parameters to their equilibrium values.  The barrier to equilibration thus arises not due to small-scale competition between the level-1 and level-2 triangles for corner tiles, but rather due to the existence of special positions and orientations of domain walls that allow the system to find deep energy minima in which different large regions of the sample break the sublattice symmetry of the level-1 structure in different ways.

%
\section{Diffraction patterns}\label{sec:diffraction}
The detection of a naturally occurring limit-periodic structure or verification of a synthetically produced one often relies on the interpretation of diffraction data.  The general features of limit-periodic diffraction have been studied by physicists and mathematicians interested in long-range aperiodic order~\cite{Godreche1989,Baake2011}, and Akiyama and Lee have proven in particular that a density pattern formed by a tiling consisting of decorated Taylor-Socolar tiles would exhibit pure point diffraction~\cite{Akiyama2012pp}.  We present here an exact calculation of the diffraction corresponding to a particular decoration of the Taylor-Socolar prototile with a density specifically designed to make the computation tractable.

Because the full limit-periodic pattern is by definition the union of a countable hierarchy of periodic patterns with increasingly larger lattice constants, we expect the diffraction pattern of any mass density associated with it to be decomposable as a sum of the form:
\begin{equation}
I(k)= \left|\sum_n \sum_{\v{b}\in B_n} N_n \hat{f_n}(k)\exp\left(i\v{u}_n\cdot \v{k}\right)\delta\left(\v{b}-\v{k}\right)\right|^2
\label{eq:gendiff}
\end{equation}
Here $B_n$ is the set of dual lattice vectors for the level-$n$ periodic pattern; $\hat{f_n}$ is the form factor of the level $n$ unit cell; the term $\exp{i\v{u}_n\cdot k}$ accounts for a potential offset of the level $n$ patterns from one another; and $N_n$ is a normalizing factor to account for the decrease in densities of the contributions with larger $n$. Note that in general $B_n\subset B_{n+1}$ because the direct lattice of the \mbox{level-$n+1$} pattern contains the basis vectors of the direct lattice of the \mbox{level-$n$} pattern.

For a Taylor-Socolar tiling, the \mbox{level-$n$} periodic pattern forms a periodic triangular lattice with lattice constant $a_n=2^n a_0$, where $a_0$ is the distance between the centers of neighboring tiles.  The $B_n$ vectors therefore form the dual triangular lattice with lattice constant $b_n=2^{-n} b_0$, and  we have $N_n=2^{-n}$.

The offsets $\v{u}_v$ specify which one of the uncountably many Socolar-Taylor tilings is under consideration.  They are expected to depend on the details of the annealing process and cannot be determined a priori; they correspond to the choice of which sublattice ($A$, $B$, $C$, or $D$) is chosen for ordering at each level, as explained in Sec.~\ref{sec:defns}.

\begin{figure}
\begin{center}
\includegraphics[width=1.0\columnwidth]{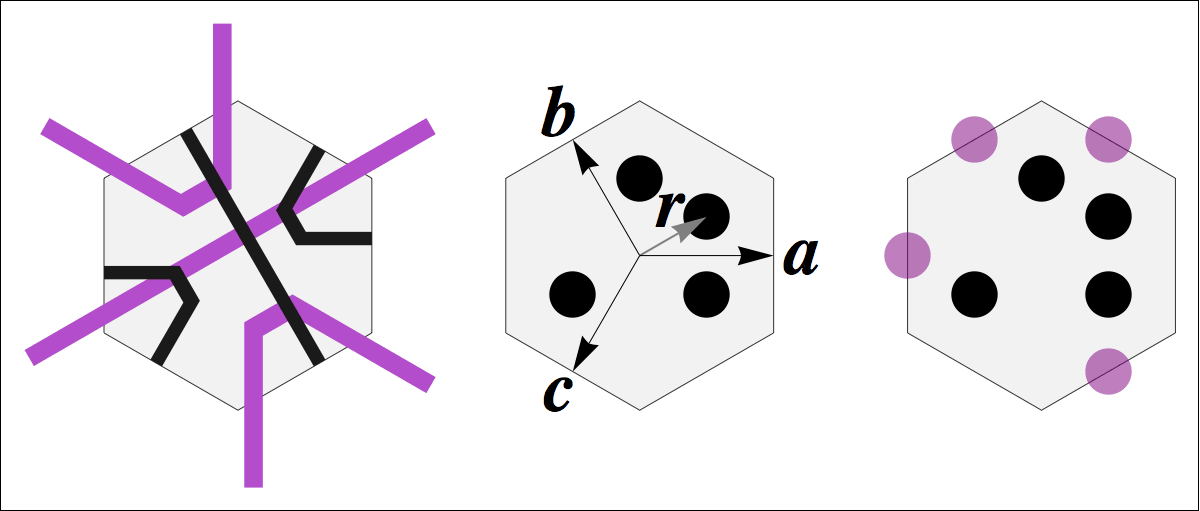}
\caption{(Color online.)  
Mass decorations used for computation of the diffraction patterns.  The tile on the right is the decoration associated with the tile orientation shown on the left.  Each disk is taken to be a point mass, all of equal weight. The black disks indicate the positions of the point masses used for the $\e_2=0$ case.
The central tile shows the vectors used in Eq.~\eqref{eq:upperdensity}.
\label{fig:masstile}}
\end{center}
\end{figure}

All that remains undetermined is the form factor of the unit cell, which depends on the particular decoration (i.e., choice of mass density) on the prototile.  To compute the form factor for the simpler $\e_2=0$ 2D tiling, we associate the prototile of Sec.~\ref{sec:eps2eq0} with a collection of four point masses arranged in the pattern shown in Fig.~\ref{fig:masstile}.
A specification of the location of these four point masses for a given tile unambiguously determines its position and orientation. Further, this decoration allows a simple calculation of the form factors for a full ground state tiling.

\begin{figure}
\begin{center}
\includegraphics[width=\columnwidth]{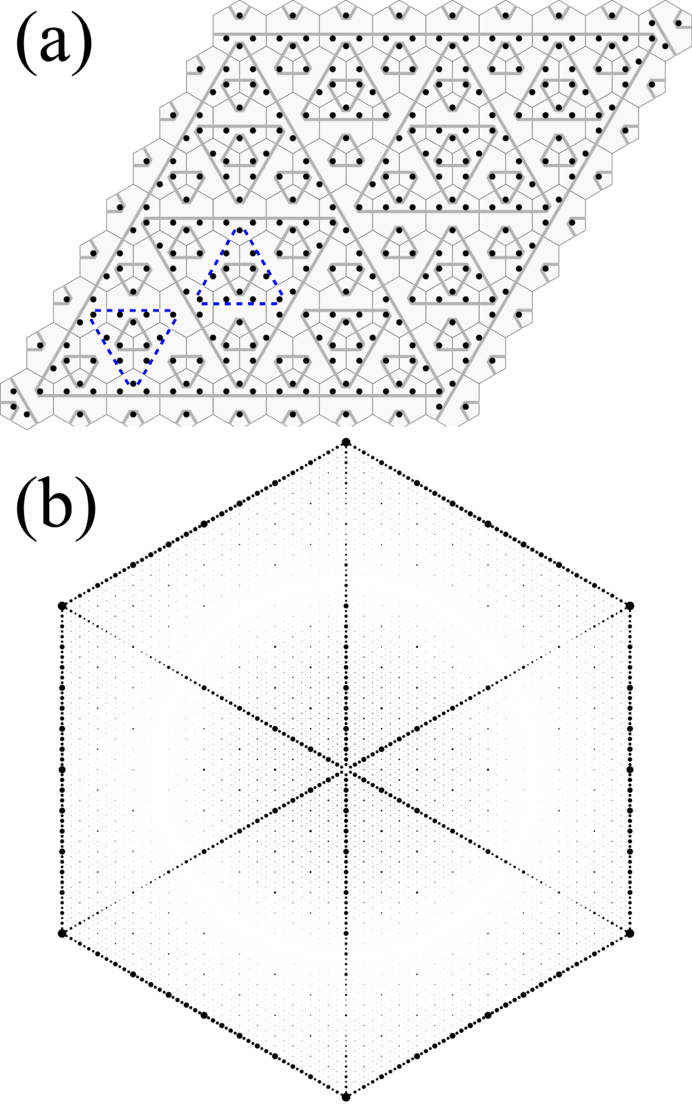} \\ 
\caption{(Color online.)
(a) The mass density for computation of the diffraction pattern for the $\e_2=0$ case (the black stripe model).  Black dots represent point masses of equal mass.  Gray lines are guides to the eye.  The point masses touching the dashed (blue) lines form the level-2 unit cell.
(b) The total computed diffraction pattern for levels 1 through 6 of the $\e_2=0$ mass decoration.  The largest wavevectors shown correspond to the basis vectors of the reciprocal lattice associated with the undecorated hexagonal tiling.
\label{fig:diffcell}}
\end{center}
\end{figure}

The calculation of the form factors and corresponding diffraction image proceeds as follows.  First, note that the inversion symmetry of the unit cell of the level-$n$ periodic pattern as shown Fig.~\ref{fig:diffcell} allows us to consider only the upper triangle of the mass decoration in our calculation.  The density of this decoration is given by equal amplitude delta functions located at each of the points shown in Fig.~\ref{fig:diffcell}.  Note that the masses shown in Fig.~\ref{fig:diffcell} are not all of the masses associated with the tiles in that figure.  Other masses on those tiles contribute to periodic structures at different levels.  The patterns formed at different levels differ only in the number of tiles inserted into the edge of each triangle, with each of those tiles contributing two masses.  The locations of the masses on the prototile have been chosen such that the spacing of masses along each triangle edge is uniform.

We define $\v{a}$, $\v{b}$, $\v{c}$ and $\v{r}$ as the constant vectors shown relative to a sample hexagon in Fig.~\ref{fig:masstile}, with the length of $\v{r}$ being half of the side length of the hexagon.  For  a general level $n$, taking the origin to be at the center of the central triangle in Fig.~\ref{fig:masstile} and defining $\kappa_n\equiv2^n-1$,
the density associated with the upper triangle is:
\begin{eqnarray}
f_{n+}(\v{x}) & = & \sum_{m=0}^{\kappa_n-1}\Big[\delta(\v{r}+m\v{a}-x)+\delta(\v{r}-m\v{c}-x)  \nonumber \\
\ & \ & \quad \quad + \;\delta(\v{r}+\kappa_n \v{a}+m \v{b} -\v{x})\Big].
\label{eq:upperdensity}
\end{eqnarray}

The Fourier transform of Eq.~\eqref{eq:upperdensity} can be written in terms of geometric sums as:
\begin{eqnarray}
\hat f_{n+}  =  e^{-i \v{r}\cdot\v{x}} \sum_{m=0}^{\kappa_n-1}\Big[ &e&\!\!^{-im\v{a}\cdot \v{x}}  +  e^{+im\v{c}\cdot \v{x}} \nonumber \\
\ &   + & \;e^{-i(\kappa_n \v{a} + m\v{b})\cdot \v{x}}\Big]
\label{eq:upperfourier}
\end{eqnarray}
Exploiting the inversion symmetry of the unit cells, the desired form factors for the periodic sub-patterns are related to Eq.~\eqref{eq:upperfourier} by
\begin{equation}
\hat f_{n}=2 \mathrm{Re}\left[\hat{f}_{n+}\right]\,.
\label{eq:formfactors}
\end{equation}

\begin{figure}
\begin{center}
\includegraphics[width=\columnwidth]{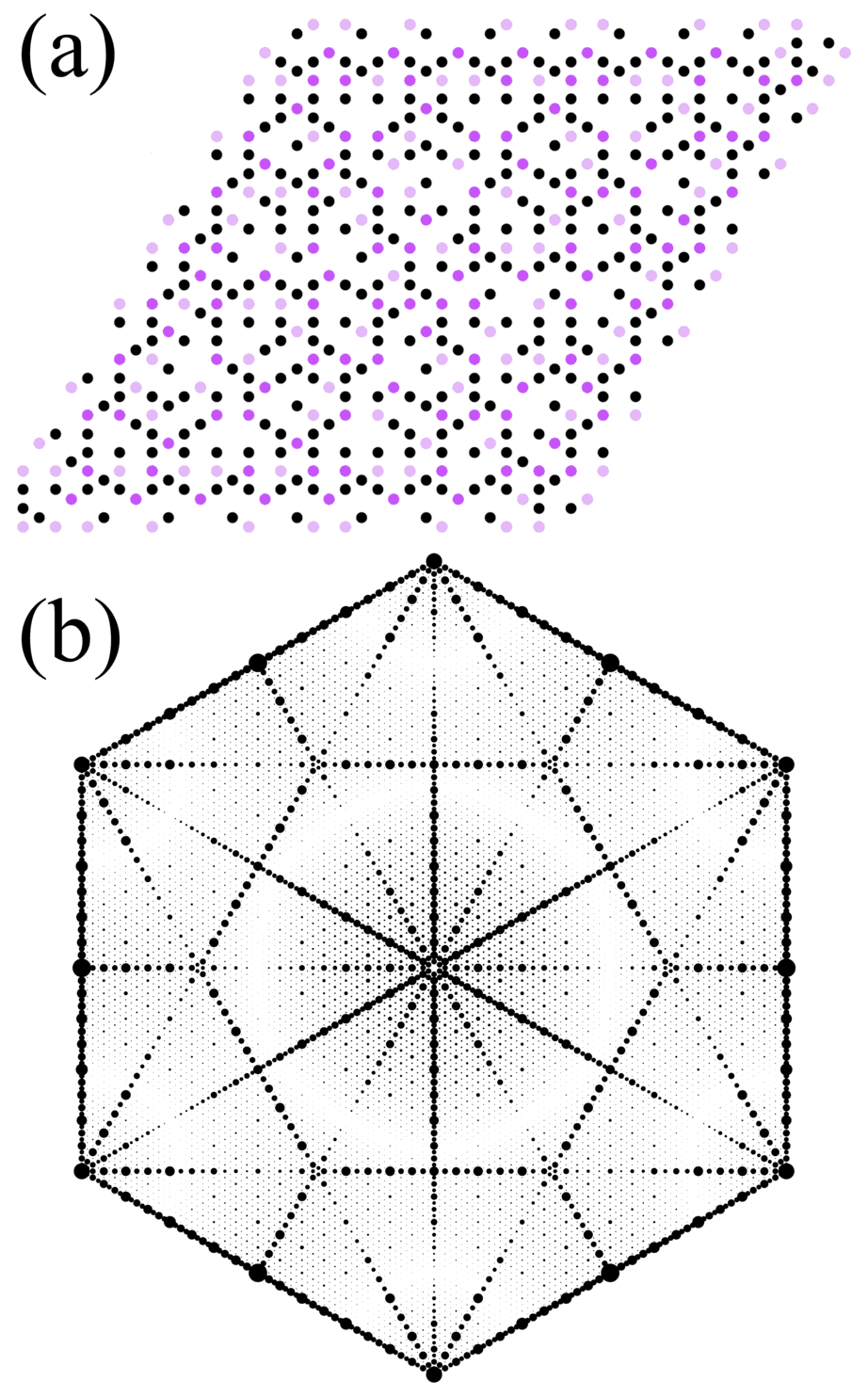} \\ 
\caption{(Color online.)  
(a) A region of the mass density for computation of the diffraction pattern for the $\e_2\neq0$ mass decoration.  Black and light gray (light purple) disks represent point particles of equal mass $m$.  Darker gray (darker purple) disks represent particles of mass $2m$.  
(b) The total computed diffraction pattern for levels 1 through 6 of the $\e_2\neq0$ mass decoration.  The largest wavevectors shown correspond to the basis vectors of the reciprocal lattice associated with the undecorated hexagonal tiling.
\label{fig:bpdensity}}
\end{center}
\end{figure}

With the form factors from Eq.~\eqref{eq:formfactors} in hand, we directly compute the diffraction patterns from this sample mass decoration. Figure~\ref{fig:diffcell} shows the diffraction image for the pattern obtained from levels 1 through 6 from Eq.~\eqref{eq:gendiff}.
Figure~\ref{fig:bpdensity} shows the diffraction image for the pattern obtained from levels 1 through 6 for the decoration associated with nonzero $\e_2$. In each of the figures, the area of each black dot is proportional to the intensity at that point on the reciprocal lattice.

The diffraction patterns illustrate the complexity of the real space structures.  The definitive feature is the lack of a smallest wavevector for an infinite sample.  We do not claim to understand which features of the global variations in intensity, such as the depleted ring at wavevectors with approximately half the magnitude of the largest wavevectors shown in these images, are generic.  The figures are presented only as illustrations of the qualitative features that might be expected if structures of this type were found in nature.

%
\section{Conclusion and remarks}
The variants of the Taylor-Socolar lattice model studied here display an intriguing array of behaviors.  In all cases, rapid quenches lead to disordered states with high barriers to equilibration.  Slow quenches, however, lead to a series of phase transitions whose limit (as $T\rightarrow 0$) is a perfect limit-periodic structure.  This is true even in cases where there exist many degenerate periodic and limit-periodic ground states.

In the 2D Taylor-Socolar model, which requires next-nearest-neighbor interactions, the limit-periodic structure is the unique ground state in the sense that any finite sample is a configuration that can be found within a single canonical instance of the Taylor-Socolar tiling.  For this model, we have studied both slow and rapid quenches.  The phase transitions are second order, and for the case $\e_2=\e_1$ a scaling theory can be used to map all of the transitions onto a single form.  We have also seen that the barriers to equilibration in rapid quenches involve particular types of domain walls that cannot move without significant increases in the energy penalty.  Finally, we have exhibited an exact diffraction pattern for two different mass decorations of the hexagonal ``unit cell.’’ 

In the 2D black stripe model, there are degenerate perfect limit-periodic structures and perfect periodic ones.  Nevertheless, upon slow cooling, the second order transitions leading to the limit-periodic structure do occur.  Roughly speaking, the highest temperature at which a periodic structure might be stable is lower than the temperature at which some level-$n$ transition occurs that creates a structure incompatible with the periodic state.

In the case of the 3D zonohedral model, we have proven that nearest-neighbor interactions are sufficient to rule out all periodic states as ground states.  With these nearest-neighbor interactions alone, there exist many degenerate limit-periodic states, having lattices of helices at any given level arranged differently relative to each other.

In the 3D cubic model, the set of degenerate ground states includes the limit-periodic states of the zonohedral model and also periodic states closely related to the periodic states of the black stripe model, and we have not ruled out the possibility of additional ground states.  Just as for the black stripe model, the limit-periodic structure does emerge during slow cooling through the same hierarchy of transitions.   In this case, however, the transitions are first order.  We have shown that the same approximate scaling relations hold here as for the Taylor-Socolar model and confirmed that they hold to high accuracy by computing free energies of the system in several phases with increasing levels of order.

The fact that the transitions appear to be second order in the zonohedral model but first order in the cubic model begs the question of how the nature of the transition changes as a function of $\e_2$.  We conjecture that the transition becomes first order for all $\e_2 < 1$, with the size of the discontinuity approaching zero as $\e_2$ approaches 1, but careful investigation of this point is beyond the scope of the present work.

One intriguing case that we have not yet studied carefully is the 3D zonohedral model with $\e_2 > \e_1$.  In the case of $\e_1 = 0$, each layer of the system forms a limit-periodic structure, but the layers are decoupled, so the structures in different layers are not likely to be in registry with each other.  For small $\e_1$, we conjecture that the level-1 transitions within each layer occur at a temperature high enough to prevent the interlayer coupling from bringing the different layers into registry, thus leading to a frustrated state at low temperature in which the black bar structures always have defects and the ground state cannot be accessed.  The resulting material would be a new type of glass whose thermodynamics and kinetics might be accessible to analysis.

Finally, the fact that the 2D black stripe model and the 3D cubic model, whose Hamiltonians involve only  relatively simple nearest-neighbor interactions, do yield limit-periodic structures upon slow cooling suggests that plausible physical interactions may indeed induce spontaneous formation of a limit-periodic structure.  The construction of a physical unit embodying these interactions could lead to a material with a thermodynamically stable structure of a type never identified previously in a spontaneously formed physical system.   The finite gap between the free energy of the relevant competing periodic phase and the partially ordered limit-periodic structure at each transition temperature implies that the path to limit-periodicity through quasistatic cooling can be followed even if next-nearest-neighbor interactions favor the periodic phase at $T=0$.  Thus the design space for physical units that might form limit-periodic phases is much larger than the strict matching rules that force the tilings may suggest.

\begin{acknowledgments}
Support for this research was provided by the NSF's Research Triangle MRSEC (DMR-1121107).  C.~M. thanks Lin Fu for useful conversations about thermodynamic integration methods.
\end{acknowledgments}
%

%
\appendix*
\section{Proof of aperiodicity of the zonohedral tile}
\begin{figure}
        \centering
        \includegraphics[width = 0.9\columnwidth]{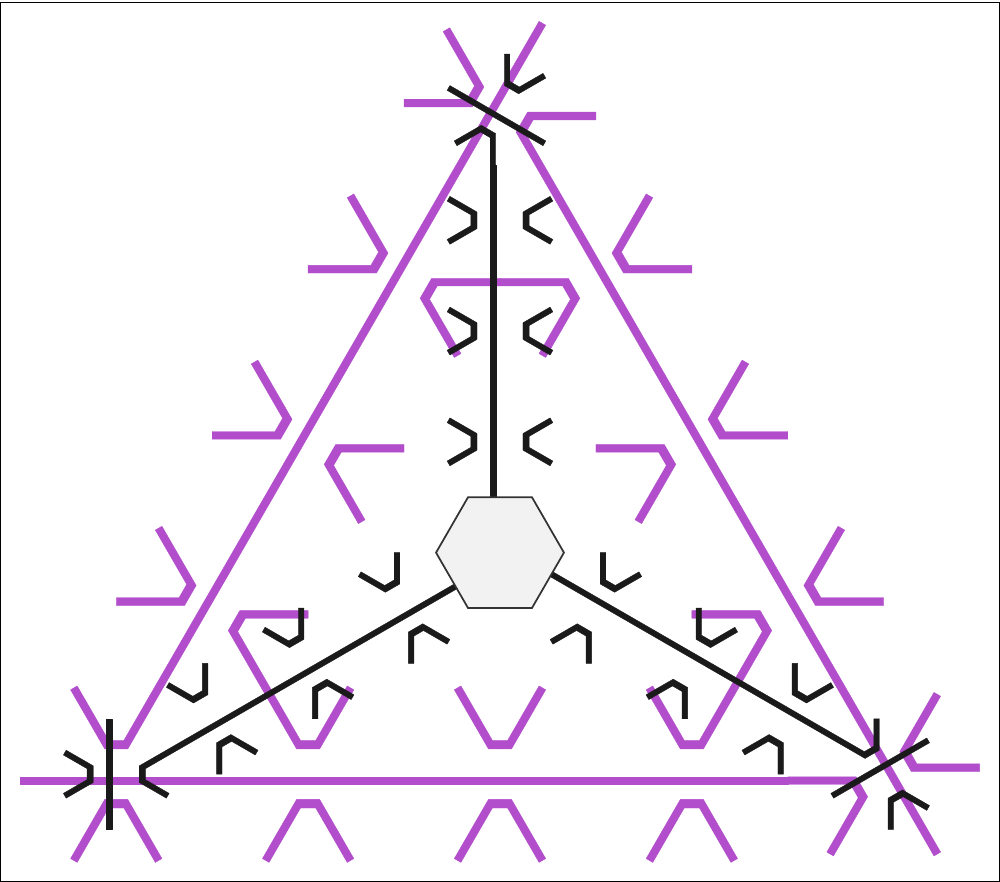}
        \caption{(Color online.)  The thick gray (purple) chiral triangle, or a smaller or larger version of it, must occur in a layer of a periodic pattern. Note that each long edge ends at a corner that turns away from another long edge. The black stripes shown are then forced as a pyramid of layers is formed. The tile at the top of the pyramid cannot match all of its black stripes to its neighbors.
        \label{fig:chiraltriangle}}
\end{figure}

We prove here that the zonohedral model, as represented by the tiel of Fig.~\ref{fig:cube}, has no periodic ground states.  The logic of the proof is as follows. First, we consider the pattern of purple stripes formed in a single layer and show that in any periodic pattern satisfying the matching rules, there must be a triangle of the type shown in Fig.~\ref{fig:chiraltriangle}.  We then show that the black stripes on this layer force the formation of a pyramid of tiles culminating in the middle with a tile that cannot match the three black stripes supporting it, shown in Fig.~\ref{fig:chiraltriangle}. Thus the purple chiral triangle is not consistent with the matching rules and no periodic tiling is possible.

In the following, we refer to a level-$n$ triangle as having side length $2^{n-1}$. Each side has $2_{n-1}-1$ tiles with straight gray (purple) stripes across them and $2$ corner tiles. Note that each tile that contributes a straight portion of the edge of a large triangle also contributes two corners of other triangles.

{\bf Lemma 1:} {\it Any closed purple triangle must be equilateral.}

{\it Proof:} All corners form angles of $\pi/3$. {\bf Q.E.D.}

\begin{figure}
        \centering
        \includegraphics[width = 0.9\columnwidth]{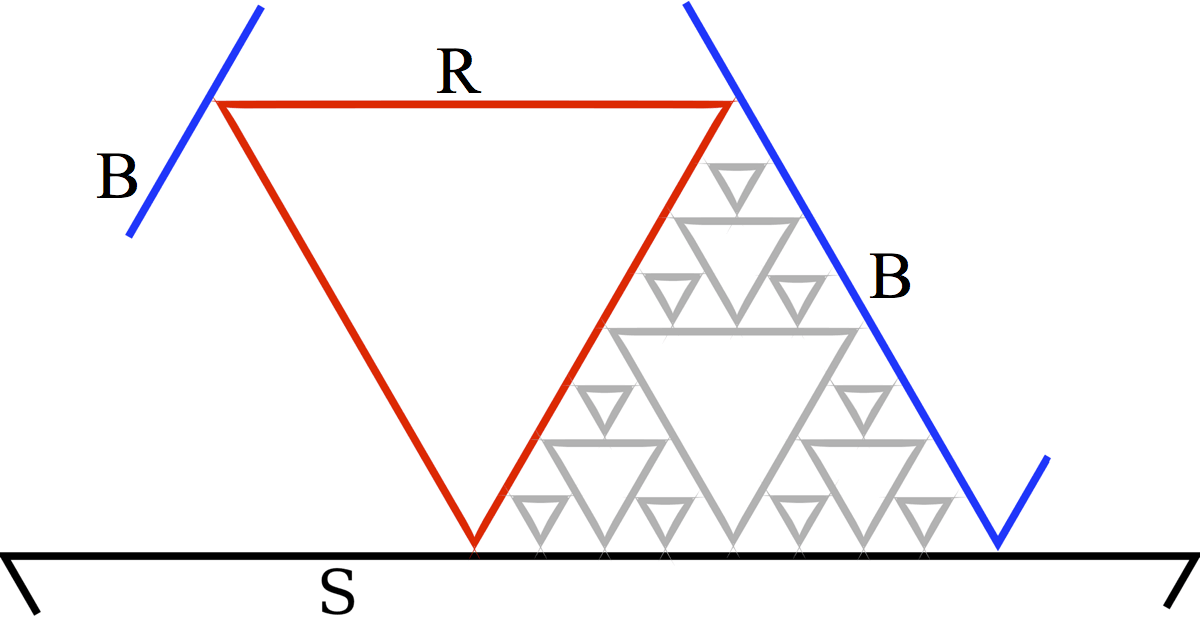}
        \caption{(Color online.)  Proof of Lemma 2.  If each edge in the red triangle (R) is shorter than one half the length of the black edge (S), then a blue edge (B) that is longer than each red one must exist.
        \label{fig:triangleedge}}
\end{figure}

{\bf Lemma 2:} {\it Let S be one side of a triangle, as shown in black in Fig.~\ref{fig:triangleedge}. At least one of the edges emanating from the corners along the exterior of the edge in question must be at least half as long as S.}

{\it Proof:} Suppose the red triangle (labeled R) shown in Fig.~\ref{fig:triangleedge} is the largest one emanating from a corner along S. The smaller triangles shown in gray are forced, as can easily be seen by inspection. If the red side length is shorter the half of S, it is clear by simple geometry that one of the blue edges must extend to meet another corner on the exterior edge of S. But this blue edge will be longer than the red one, which contradicts the claim that the red one was the largest. {\bf Q.E.D.}
\begin{figure}
        \centering
        \includegraphics[width =0.9\columnwidth]{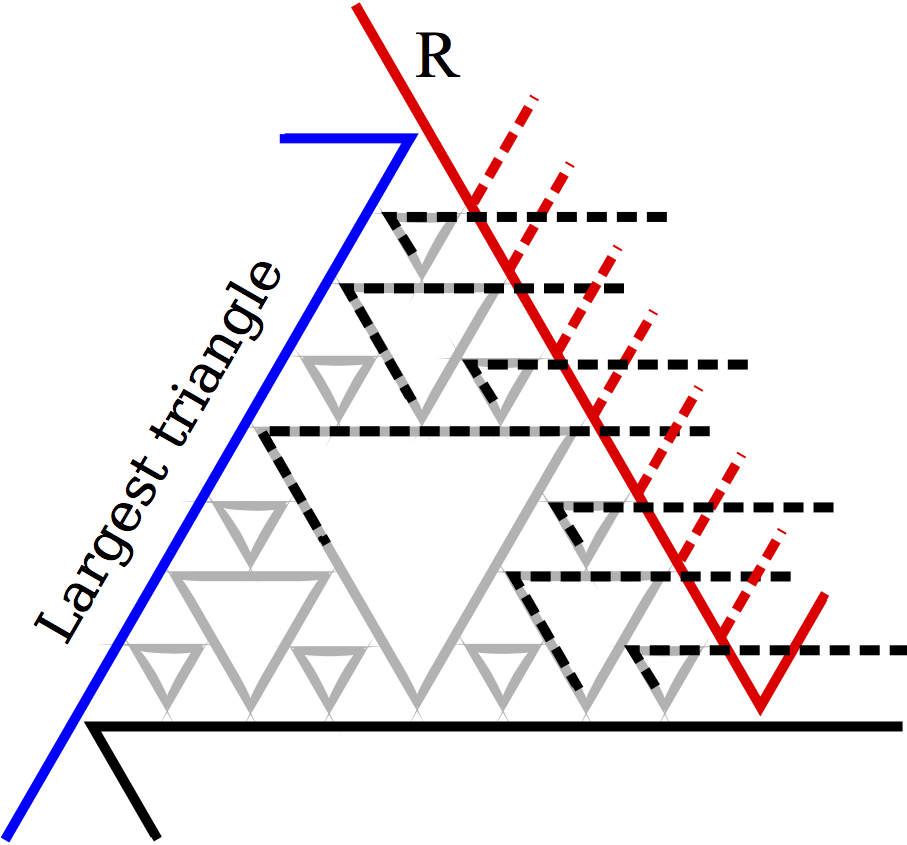}
        \caption{(Color online.)  Proof of Lemma 4.  All possible choices of red lines (R) and corresponding black lines lead to the formation of a chiral triangle of the type shown in Fig.~\ref{fig:chiraltriangle}.
        \label{fig:triangleregion}}
\end{figure}

{\bf Lemma 3:} {\it There can be no infinite line in a periodic tiling that satisfies the purple stripe rules everywhere.}

{\it Proof:} If there is an infinite line, then by Lemma 2, there would have to be infinite half-lines in each of the three triangular directions. Given that lines cannot intersect, this rules out any periodic structure. {\bf Q.E.D.}

{\bf Corrollary:} {\it There must be a largest triangle in the periodic structure.} (The proof is obvious.)

{\bf Lemma 4:} {\it In any periodic structure satisfying the purple stripe matching rules, there must exist an equilateral triangle bounded by the edges of three separate triangles. Furthermore, the orientations of the tiles at the corners of this triangle form a chiral structure as depicted in Fig.~\ref{fig:triangleregion}.}

{\it Proof:} In Fig.~\ref{fig:triangleregion}, let the blue line (labeled ``Largest triangle'') be a portion of one of the largest triangles in the structure.  At the blue corner, there must be an edge directed as shown in red. Because the red edge (labeled R) cannot be longer than the blue one, the red corner shown must turn away from the blue edge, as shown, otherwise it would be impossible to complete the red triangle. Working from the top tile where the red line meets the blue corner, the gray triangles are forced and the red corner must occur at one of the places indicated by the red dashed segments. Where the red corner occurs, there must be a horizontal stripe as shown in black, and the black edge must terminate in a corner that turns downward as shown. By inspection, every possible choice for the black edge results in the formation of an equilateral triangular region consisting of the edges of three separate triangles: one red, one black, and the other either gray or blue. {\bf Q.E.D.}

We have thus shown that any given layer of any periodic tiling obeying the purple stripe rules everywhere must contain a triangular region of the type shown in purple in Fig.~\ref{fig:chiraltriangle}. We now consider the layer above this ``chiral triangle,'' which is coupled to it through the matching of black bars. We wish to show that the pattern of black bars in Fig.~\ref{fig:chiraltriangle} is forced.

{\bf Lemma 5:} {\it The black bars passing through the corners of a chiral triangle must extend all the way to a single tile directly above (or below) the center of the triangle.}

\begin{figure}
        \centering
        \includegraphics[width = 0.9\columnwidth]{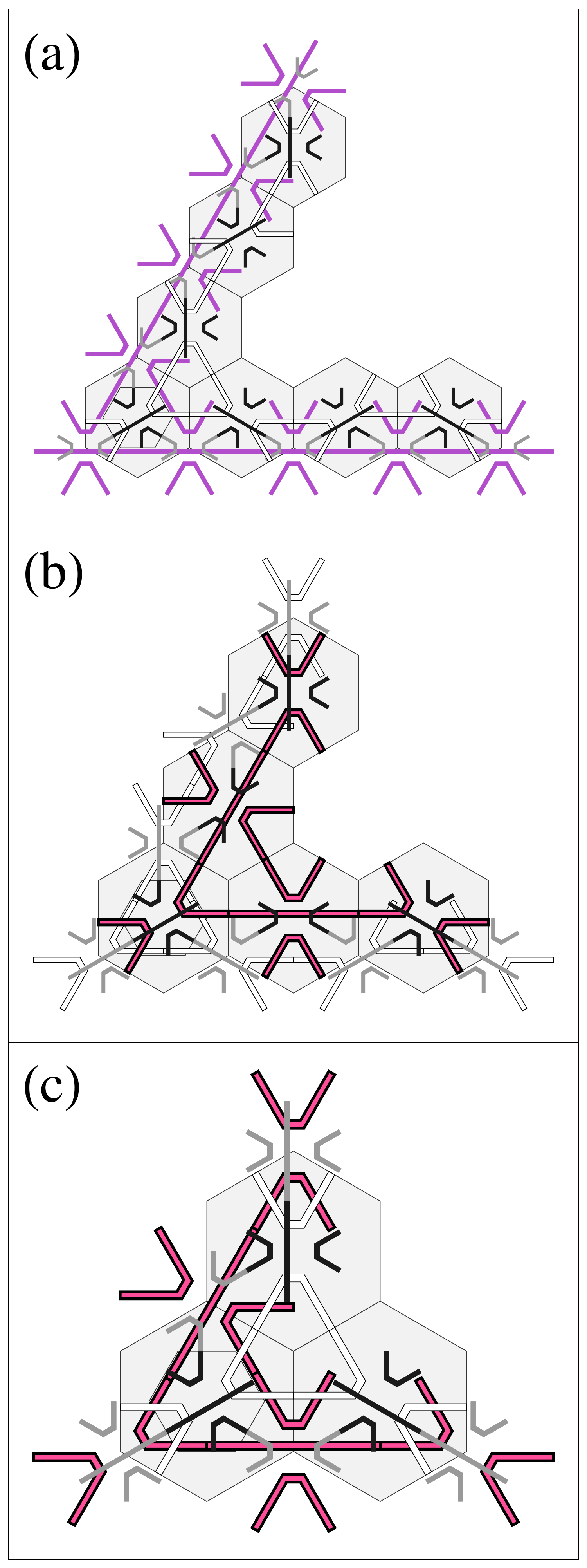}
        \caption{(Color online.)  Proof of Lemma 5. (a) Base layer (thick gray, (purple)) required by Lemma 5 and forced layer (outlined white) above it.  (b) Outlined white layer of (a) and forced layer (outlined gray (red)) above it.  (c)  Outlined gray (red) layer of (b) and forced layer (outlined white) above it. 
        \label{fig:chiralproof}}
\end{figure}

{\it Proof:} The proof is illustrated in Fig.~\ref{fig:chiralproof}.
Panel (a) of the figure shows one corner of a large chiral triangle in thick gray (purple). The black bars associated with the tiles in the layer of thick gray (purple) stripes are shown in thin gray. Only the portions of the black bars whose locations are forced are shown. The thick gray (purple) layer forces the placement of some of the tiles one layer above it, and these tiles are shown in outlined white, with the forced portions of their black bar decorations shown in black. The key feature is that the long black stripe on the tile at the lower left corner is forced. Similarly, panel (b) shows the forced tile decorations on the next layer up, shown in outlined gray (red), and panel (c) shows one layer above that, again in outlined white. Note that the two outlined white layers have the same structure, implying that the pattern must repeat and the black bars must extend upwards as shown in Fig.~\ref{fig:chiraltriangle}. {\bf Q.E.D.}

The proof of aperiodicity is now complete, for Lemma 5 guarantees the existence of a tile location in any periodic pattern (the tile at the central site of Fig.~\ref{fig:chiraltriangle}) for which there is no way to place a tile that satisfies the black bar matching rule.
\clearpage

\end{document}